\documentclass[twocolumn]{aastex631}
\usepackage{amssymb}

\newcommand\event{GRB\,221009A}
\newcommand\swift{Swift}
\newcommand\nicer{NICER}
\newcommand\maxi{MAXI}
\newcommand\fermi{Fermi}
\def\arcsec{\ensuremath{^{\prime\prime}}}
\def\arcmin{\ensuremath{^\prime}}
\def\cms{\ensuremath{$cm$^{-2}}}
\newcommand{\tim}[1]{\ensuremath{\times 10^{#1}}}
\newcommand{\tz}{$T_0$}
\newcommand{\tzks}[1]{$T_0{#1}$\,ks}
\def\nh{\ensuremath{N_{\rm H}}}
\def\ecs{\ensuremath{$erg~cm$^{-2}~$s$^{-1}}}

\begin{document}

\title{\event: Discovery of an Exceptionally Rare Nearby and Energetic Gamma-Ray Burst}

\author[0000-0002-0025-3601]{Maia A. Williams}
\affiliation{Department of Astronomy and Astrophysics, Pennsylvania State University, 525 Davey Lab, University Park, PA 16802, USA}

\author[0000-0002-6745-4790]{Jamie A. Kennea}
\affiliation{Department of Astronomy and Astrophysics, Pennsylvania State University, 525 Davey Lab, University Park, PA 16802, USA}

\author[0000-0001-6849-1270]{S. Dichiara}
\affiliation{Department of Astronomy and Astrophysics, Pennsylvania State University, 525 Davey Lab, University Park, PA 16802, USA}

\author{Kohei Kobayashi}
\affiliation{Department of Physics, Nihon University, 1-8 Kanda-Surugadai, Chiyoda-ku, 
Tokyo 101-8308, Japan}

\author[0000-0002-0207-9010]{Wataru B. Iwakiri} 
\affiliation{International Center for Hadron Astrophysics, Chiba University, Chiba 263-8522, Japan} 

\author[0000-0001-5186-5950]{Andrew P. Beardmore}
\affiliation{School of Physics and Astronomy, University of Leicester, University Road, Leicester, LE1 7RH, UK}

\author[0000-0002-8465-3353]{P.A. Evans}
\affiliation{School of Physics and Astronomy, University of Leicester, University Road, Leicester, LE1 7RH, UK}

\author[0000-0002-8433-8652]{Sebastian Heinz}
\affiliation{Department of Astronomy, University of Wisconsin-Madison, 475 N Charter St, Madison, WI 53726, USA}

\author[0000-0002-7851-9756]{Amy Lien}
\affiliation{University of Tampa, Department of Chemistry, Biochemistry, and Physics, 401 W. Kennedy Blvd, Tampa, FL 33606, USA}

\author[0000-0001-9309-7873]{S.~R.~Oates}
\affiliation{School of Physics and Astronomy and Institute for Gravitational Wave Astronomy, University of Birmingham, B15 2TT, UK}

\author[0000-0003-0939-1178]{Hitoshi Negoro}
\affiliation{Department of Physics, Nihon University, 1-8 Kanda-Surugadai, Chiyoda-ku, 
Tokyo 101-8308, Japan}

\author[0000-0003-1673-970X]{S. Bradley Cenko}
\affiliation{Astrophysics Science Division, NASA Goddard Space Flight Center, 8800 Greenbelt Road, Greenbelt, MD 20771, USA}
\affiliation{Joint Space-Science Institute, University of Maryland, College Park, MD 20742, USA}

\author[0000-0002-5341-6929]{Douglas J. K. Buisson} 
\affiliation{independent researcher} 

\author[0000-0002-8028-0991]{Dieter H. Hartmann}
\affiliation{Department of Physics and Astronomy, Clemson University, Clemson, SC 29634, USA}

\author[0000-0002-6789-2723]{Gaurava K. Jaisawal} 
\affiliation{DTU Space, Technical University of Denmark, Elektrovej 327-328, DK-2800 Lyngby, Denmark}

\author[0000-0003-4650-4186]{N.P.M. Kuin}
\affiliation{UCL, Space and Climate Physics/MSSL, Holmbury St. Mary, Dorking, Surrey RH5 6NT United Kingdom}

\author[0000-0001-8058-9684]{Stephen Lesage}
\affiliation{Department of Space Science, University of Alabama in Huntsville, 320 Sparkman Drive, Huntsville, AL 35899, USA}
\affiliation{Center for Space Plasma and Aeronomic Research, University of Alabama in Huntsville, Huntsville, AL 35899, USA}

\author[0000-0001-5624-2613]{Kim L. Page}
\affiliation{School of Physics and Astronomy, University of Leicester, University Road, Leicester, LE1 7RH, UK}

\author[0000-0002-4299-2517]{Tyler Parsotan}
\affiliation{Center for Space Science and Technology, University of Maryland Baltimore County, 1000 Hilltop Circle, Baltimore, MD 21250, USA}
\affiliation{Astrophysics Science Division, NASA Goddard Space Flight Center, 8800 Greenbelt Road, Greenbelt, MD 20771, USA}
\affiliation{Center for Research and Exploration in Space Science and Technology, NASA/GSFC, Greenbelt, Maryland 20771, USA}

\author[0000-0003-1386-7861]{Dheeraj R. Pasham} 
\affiliation{MIT Kavli Institute for Astrophysics and Space Research} 

\author[0000-0001-6620-8347]{B. Sbarufatti}
\affiliation{INAF - Osservatorio Astronomico di Brera, Via Bianchi 46, 23807 Merate (LC), Italy}

\author[0000-0003-1817-3009]{Michael H. Siegel}
\affiliation{Department of Astronomy and Astrophysics, Pennsylvania State University, 525 Davey Lab, University Park, PA 16802, USA}

\author{Satoshi Sugita} 
\affiliation{Department of Physical Science, Aoyama Gakuin University, 
5-10-1 Fuchinobe, Chuo-ku, Sagamihara, Kanagawa 252-5258, Japan} 

\author[0000-0002-7991-028X]{George Younes} 
\affiliation{Astrophysics Science Division, NASA Goddard Space Flight Center, 8800 Greenbelt Road, Greenbelt, MD 20771, USA} 
\affiliation{Department of Physics, The George Washington University, 725 21st St. NW, Washington, DC 20052, USA} 

\author[0000-0002-9731-8300]{Elena Ambrosi}
\affiliation{INAF - Istituto di Astrofisica Spaziale e Fisica Cosmica di Palermo, Via U. La Malfa 153, I-90146 Palermo, Italy}

\author{Zaven Arzoumanian} 
\affiliation{Astrophysics Science Division, NASA Goddard Space Flight Center, 8800 Greenbelt Road, Greenbelt, MD 20771, USA} 

\author[0000-0001-6106-3046]{M. G. Bernardini}
\affiliation{INAF - Osservatorio Astronomico di Brera, Via Bianchi 46, 23807
Merate (LC), Italy}

\author[0000-0001-6278-1576]{S. Campana}
\affiliation{INAF - Osservatorio Astronomico di Brera, Via Bianchi 46, 23807 Merate (LC), Italy}

\author[0000-0002-9558-2394]{Milvia Capalbi}
\affiliation{INAF - Istituto di Astrofisica Spaziale e Fisica Cosmica di Palermo, Via U. La Malfa 153, I-90146 Palermo, Italy}

\author[0000-0002-9280-836X]{Regina Caputo}
\affiliation{Astrophysics Science Division, NASA Goddard Space Flight Center, 8800 Greenbelt Road, Greenbelt, MD 20771, USA}
\affiliation{Joint Space-Science Institute, University of Maryland, College Park, MD 20742, USA}

\author[0000-0002-5042-1036]{Antonino D'A\`{i}}
\affiliation{INAF - Istituto di Astrofisica Spaziale e Fisica Cosmica di Palermo, Via U. La Malfa 153, I-90146 Palermo, Italy}

\author[0000-0001-7164-1508]{P. D’Avanzo}
\affiliation{INAF - Osservatorio Astronomico di Brera, Via Bianchi 46, 23807
Merate (LC), Italy}

\author[0000-0002-7320-5862]{V. D’Elia}
\affiliation{ASI - Italian Space Agency, Space Science Data Centre, Via del Politecnico snc, 00133 Rome, Italy}

\author[0000-0002-4036-7419]{Massimiliano De Pasquale}
\affiliation{Department of Mathematics, Informatics, Physics, Earth Sciences of University of Messina, Polo Papardo, Via F. D’Alcontres 31, 98165 Messina, Italy}

\author[0000-0002-8775-2365]{R. A. J. Eyles-Ferris}
\affiliation{School of Physics and Astronomy, University of Leicester, University Road, Leicester, LE1 7RH, UK}

\author[0000-0001-7828-7708]{Elizabeth Ferrara} 
\affil{Department of Astronomy, University of Maryland, College Park, MD, 20742, USA}
\affil{Center for Research and Exploration in Space Science and Technology, NASA/GSFC, Greenbelt, Maryland 20771, USA}
\affil{Astrophysics Science Division, NASA Goddard Space Flight Center, 8800 Greenbelt Road, Greenbelt, MD 20771, USA}

\author[0000-0001-7115-2819]{Keith C. Gendreau} 
\affiliation{Astrophysics Science Division, NASA Goddard Space Flight Center, 8800 Greenbelt Road, Greenbelt, MD 20771, USA} 

\author[0000-0002-9090-5553]{Jeffrey D. Gropp}
\affiliation{Department of Astronomy and Astrophysics, Pennsylvania State University, 525 Davey Lab, University Park, PA 16802, USA}

\author[0000-0001-9656-0261]{Nobuyuki Kawai} 
\affiliation{Department of Physics, Tokyo Institute of Technology, 2-12-1 Ookayama, 
Meguro-ku, Tokyo 152-8551, Japan} 

\author[0000-0002-7465-0941]{Noel Klingler}
\affiliation{Center for Space Science and Technology, University of Maryland Baltimore County, 1000 Hilltop Circle, Baltimore, MD 21250, USA}
\affiliation{Astrophysics Science Division, NASA Goddard Space Flight Center, 8800 Greenbelt Road, Greenbelt, MD 20771, USA}
\affiliation{Center for Research and Exploration in Space Science and Technology, NASA/GSFC, Greenbelt, Maryland 20771, USA}

\author[0000-0003-2714-0487]{Sibasish Laha} 
\affiliation{Center for Space Science and Technology, University of Maryland Baltimore County, 1000 Hilltop Circle, Baltimore, MD 21250, USA}
\affiliation{Astrophysics Science Division, NASA Goddard Space Flight Center, 8800 Greenbelt Road, Greenbelt, MD 20771, USA}
\affiliation{Center for Research and Exploration in Space Science and Technology, NASA/GSFC, Greenbelt, Maryland 20771, USA}

\author[0000-0002-2810-2143]{A. Melandri}
\affiliation{INAF - Osservatorio Astronomico di Roma, Via Frascati 33, 00040
Monte Porzio Catone (RM), Italy}

\author[0000-0002-6337-7943]{Tatehiro Mihara} 
\affiliation{High energy astrophysics, RIKEN, Wako, Saitama, 351-0198, Japan} 

\author[0000-0002-1103-7082]{Michael Moss} 
\affiliation{Department of Physics, The George Washington University, 725 21st St. NW, Washington, DC 20052, USA} 

\author[0000-0002-5218-1899]{Paul O’Brien}
\affiliation{School of Physics and Astronomy, University of Leicester, University Road, Leicester, LE1 7RH, UK}

\author[0000-0001-5624-2613]{Julian P. Osborne}
\affiliation{School of Physics and Astronomy, University of Leicester, University Road, Leicester, LE1 7RH, UK}

\author[0000-0001-7128-0802]{David M. Palmer}
\affiliation{Los Alamos National Laboratory, Los Alamos, NM, 87545, USA}

\author[0000-0003-3613-4409]{Matteo Perri}
\affiliation{INAF - Osservatorio Astronomico di Roma, Via Frascati 33, 00040
Monte Porzio Catone (RM), Italy}
\affiliation{ASI - Italian Space Agency, Space Science Data Centre, Via del Politecnico snc, 00133 Rome, Italy}

\author{Motoko Serino} 
\affiliation{Department of Physical Science, Aoyama Gakuin University, 
5-10-1 Fuchinobe, Chuo-ku, Sagamihara, Kanagawa 252-5258, Japan} 

\author{E.~Sonbas}
\affiliation{Adiyaman University, Department of Physics, 02040 Adiyaman, Turkey}

\author[0000-0001-7253-8553]{Michael Stamatikos}
\affiliation{Department of Physics, The Ohio State University, 191 West Woodruff Avenue, Columbus, OH 43210, USA}
\affiliation{Department of Astronomy, The Ohio State University, Columbus, OH 43210, USA}
\affiliation{Center for Cosmology and AstroParticle Physics, The Ohio State University, 191 West Woodruff Avenue, Columbus, OH 43210, USA}

\author[0000-0001-5803-2038]{Rhaana Starling}
\affiliation{School of Physics and Astronomy, University of Leicester, University Road, Leicester, LE1 7RH, UK}

\author[0000-0003-0121-0723]{G. Tagliaferri}
\affiliation{INAF - Osservatorio Astronomico di Brera, Via Bianchi 46, 23807
Merate (LC), Italy}

\author[0000-0002-2810-8764]{Aaron Tohuvavohu}
\affiliation{David A. Dunlap Department of Astronomy \& Astrophysics, University of Toronto, 50 St. George Street, Toronto, ON, M5S3H4 Canada}

\author[0000-0001-5326-880X]{Silvia Zane}
\affiliation{Mullard Space Science Laboratory, University College London, Holmbury St Mary, Dorking, Surrey, RH5 6NT, UK}

\author{Houri Ziaeepour}
\affiliation{Observatoire de Besan\c{c}on, Universit\'e de Franche Compt\'e, 41 bis ave. de l'Observatoire, BP 1615, 25010 Besan\c{c}on, France}
\affiliation{Mullard Space Science Laboratory, University College London, Holmbury St Mary, Dorking, Surrey, RH5 6NT, UK}

\begin{abstract}
We report the discovery of the unusually bright long-duration gamma-ray burst (GRB), \event{}, as observed by 
the Neil Gehrels Swift Observatory (\swift), Monitor of All-sky X-ray Image (\maxi{}), and Neutron Star Interior Composition Explorer Mission (\nicer{}). This energetic GRB was located relatively nearby ($z = 0.151$), allowing for sustained observations of the afterglow. The large X-ray luminosity and low Galactic latitude ($b = 4.3^{\circ}$) make \event\ a powerful probe of dust in the Milky Way. Using echo tomography we map the line-of-sight dust distribution and find evidence for significant column densities at large distances ($\gtrsim 10$\,kpc). We present analysis of the light curves and spectra at X-ray and UV/optical wavelengths, and find that the X-ray afterglow of \event\ is more than an order of magnitude brighter at $T_{0}+4.5$\,ks than any previous GRB observed by \swift. In its rest frame \event\ is at the high end of the afterglow luminosity distribution, but not uniquely so. In a simulation of randomly generated bursts, only 1 in 10$^4$ long GRBs were as energetic as \event; such a large $E_{\gamma,\mathrm{iso}}$ implies a narrow jet structure, but the afterglow light curve is inconsistent with simple top-hat jet models. Using the sample of \swift\ GRBs with redshifts, we estimate that GRBs as energetic and nearby as \event\ occur at a rate of $\lesssim$ 1 per 1000\,yr -- making this a truly remarkable opportunity unlikely to be repeated in our lifetime. 
\end{abstract}


\section{Introduction} \label{sec:intro}

Massive stars exhibit a broad continuum of properties in their terminal explosions. At one end are the cosmological long-duration gamma-ray bursts (GRBs), capable of coupling tremendous energies ($\gtrsim 10^{51}$\,erg) to highly collimated ejecta with a bulk Lorentz factor $\Gamma_{0} \gtrsim 100$ \citep{Piran2004}. On the other end, the energy budget of most stripped-envelope core-collapse supernovae is dominated by the (quasi)-isotropic supernova emission, with photospheric velocities of tens of thousands of km\,s$^{-1}$ (e.g., \citealt{Liu+2016}). Intermediate between these two extremes lies the growing class of low-luminosity GRBs and relativistic supernovae (e.g., \citealt{Margutti+2014}). Typified by the prototypical low-luminosity GRB 980425 associated with SN\,1998bw
\citep{Galama+1998,Kulkarni+1998}, these sources couple several orders of magnitude less energy to their moderately relativistic ejecta ($\sim 10^{48}$\,erg), and lack the high degree of collimation of cosmological GRBs. 

In the nearby universe ($z \lesssim 0.3$), where high-energy facilities are sensitive to low-luminosity GRBs, high-luminosity GRBs are exceedingly rare due to their much lower volumetric rate. Yet an energetic GRB observed within this volume could produce unprecedented brightness. Here we report the discovery of \event, an extremely luminous GRB \citep{GCN:32762} in our cosmic backyard.
 
On 2022 October 9 at 14:10:17~UT, the Burst Alert Telescope (BAT; \citealt{Barthelmy+2005}), onboard the Neil Gehrels Swift Observatory (\swift{}; \citealt{Gehrels+2004}), triggered twice in rapid succession on a new cosmic source in the constellation Sagitta. Following its automated burst response, \swift{} promptly slewed to the location of the first trigger, detecting a bright transient seen with both the \swift\ X-Ray Telescope (XRT; \citealt{BurrowsXRT}) and the Ultraviolet/Optical Telescope (UVOT; \citealt{UVOT}). 
Due to the rarity of such a repeated BAT image trigger and the proximity of the source to the Galactic plane ($b$ = 4.3$^{\circ}$), it was initially classified as a new Galactic X-ray and optical transient, and therefore was designated Swift\,J1931.1+1946 \citep{GCN:32632}. 
Monitor of All-sky X-ray Image (\maxi{}; \citealt{matsuoka09}) reported the detection of bright X-ray emission from this location shortly thereafter \citep{ngrgrb}.

Subsequently the Gamma-ray Burst Monitor (GBM; \citealt{Meegan+2009}) onboard \fermi{} reported the detection of an exceptionally bright long-duration GRB $\approx$55\,min prior to the initial BAT trigger, with a consistent localization \citep{GCN:32636}. Due to issues in receiving data, the automated classification and localization notices associated with this onboard GBM trigger were not distributed to the world. However, the \fermi{} team rapidly communicated the existence of the GBM trigger, and its spatial coincidence with the double BAT trigger, to the \swift{} team. This spatial association, together with analysis of prompt XRT data that showed a smooth GRB-like power-law decline, led to the conclusion that Swift\,J1931.1+1946 was in fact a GRB, \event\ \citep{GCN:32635}. For the first time in the $\sim$18 years since launch, BAT triggered not on the GRB prompt emission, but instead on the bright high-energy afterglow when \event{} entered the field of view. 

The unusual brightness of \event\ prompted widespread follow-up at multiple wavelengths. Additional X-ray detections were reported by the Neutron Star Interior Composition Explorer (\nicer{}; \citealt{GCN:32694}) and the Nuclear Spectroscopic Telescope Array (NuSTAR; \citealt{GCN:32695}). The Large High Altitude Air Shower Observatory reported detecting photons up to 18 TeV \citep{GCN:32677}. Changes in the strength of signals propagated by radio transmitters were recorded at the time of the GBM trigger as the photons from the GRB ionized Earth’s atmosphere \citep{GCN:32744,GCN:32745}. Spectroscopic observations of the afterglow and the host galaxy provided a redshift of $z = 0.151$ \citep{GCN:32648, GCN:32686, GCN:32765}, corresponding to a distance of 749.3 Mpc.

This paper is organized as follows: $\S$2 contains analysis of the observations taken by \swift\ (BAT, XRT, and UVOT), \maxi{}, and \nicer{}; in $\S$3, we present analysis of the dust scattering echo, broadband spectrum, and light curve of the burst afterglow; we discuss how \event\ compares to other GRBs in $\S$4, investigate the astrophysical rate of similar events, and the nature of energetic GRBs; in $\S$5, we present our conclusions. We show that due to the combination of proximity and large (but not unprecedented) intrinsic luminosity \event{} has a much brighter X-ray afterglow than previously observed \swift{} GRBs, and such luminous nearby events are extremely rare occurrences. 

For this paper, we assume a cosmology with $H_0 = 67.36$, $\Omega_m = 0.3153$, $\Omega_v= 0.6847$ \citep{Planck+2020}, and $z = 0.151$ unless otherwise stated. We adopt the GBM trigger time as the burst onset, i.e., $T_{0}$ = 13:16:59.99~UTC on 2022 October 9. Magnitudes are reported on the Vega system, and uncertainties are given at a 90\% confidence interval (unless otherwise noted).   

\section{Observations} \label{sec:obs}

\subsection{\swift\ Burst Alert Telescope} \label{sec:bat}
\begin{figure*}
    \centering
    \includegraphics[width=\linewidth]{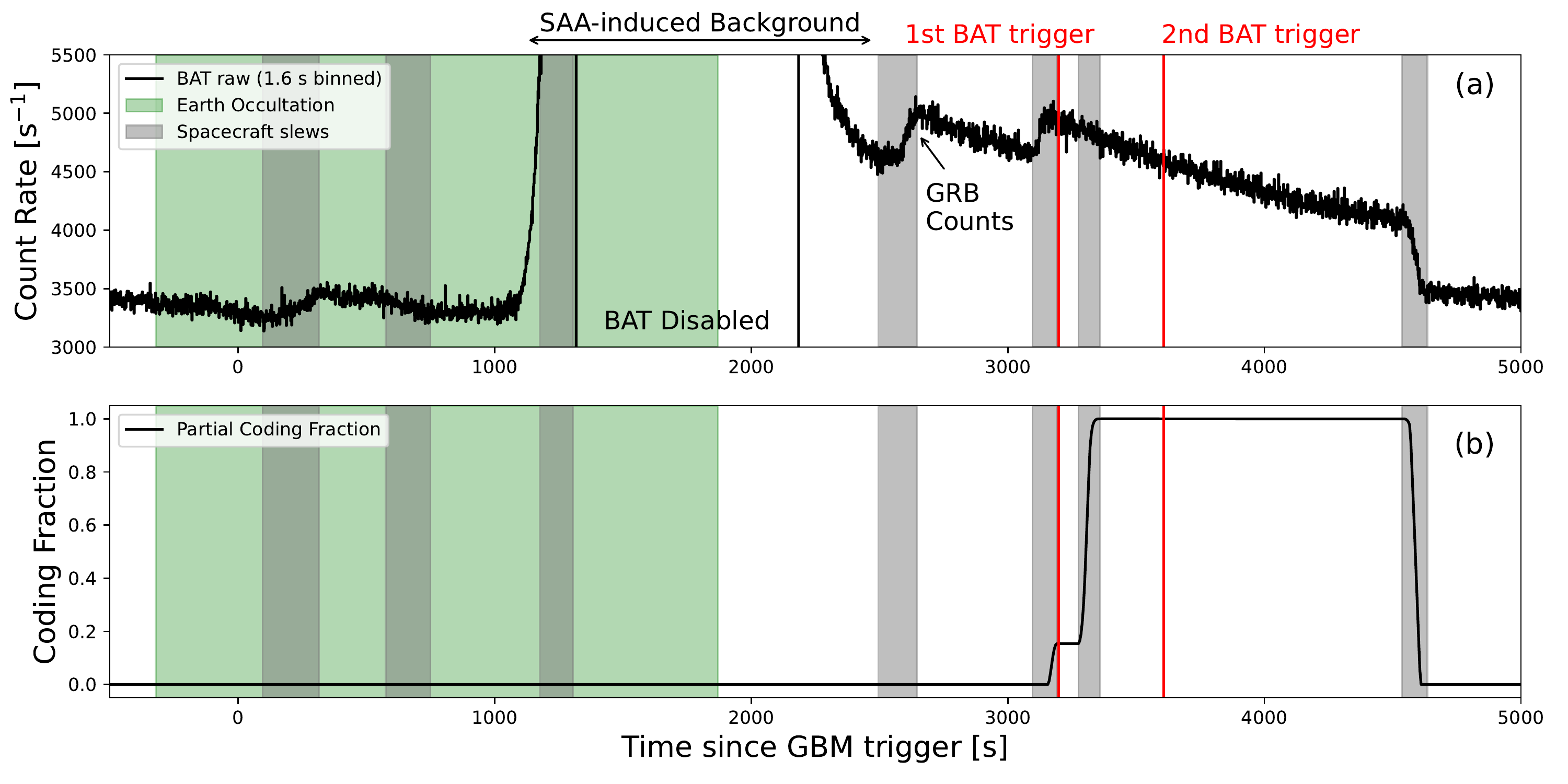}
    \caption{BAT raw light curve from the rate data. Panel (a) shows the raw light curve in the 15--350\,keV band with a time binning of 1.6\,s. This light curve was made from BAT quad-rate data, which records continuous count rates from all active detectors in four different energy ranges (15--25, 25--50, 50--100, and 100--350\,keV). The time period when the GRB was occulted by the Earth (within 69$^{\circ}$ of the Earth center) is marked in green. Spacecraft slew times are marked in gray. The two BAT trigger times are marked by red lines. Panel (b) shows the partial coding fraction as a function of time. A partial coding fraction of zero indicates that the GRB was outside of the BAT coded field of view, and a value of one indicates that the source is in the highest sensitivity region of the BAT coded field of view.}
    \label{fig:rate_lc}
\end{figure*}

Figure~\ref{fig:rate_lc} shows the BAT raw light curves (i.e., not background-subtracted), summed over all detectors, from $T_{0}-500$\,s to $T_{0}+5000$\,s. The location of \event\ was occulted by the Earth until $T_{0}+1870$\,s. At $\sim$$T_{0}+1100$\,s, the overall count rate began to rise due to increased particle background as \swift{} approached the South Atlantic Anomaly (SAA). From $T_{0}+$1317--2183\,s, BAT data collection was disabled as \swift{} transited the SAA. The count rate remained elevated while exiting the SAA until the spacecraft slew beginning at $T_{0}+2496$\,s, when we attribute the linearly decaying enhanced count rate to emission from \event{}. However, the source location was not in the coded portion of the BAT field of view until a slew beginning at $T_{0}+3095$\,s. 

Finally after this slew completed, BAT triggered on \event, at 14:10:18 ($T_{0}+3199$\,s; trigger ID 1126853) and 14:17:06 ($T_{0}+3607$\,s; trigger ID 1126854) UTC. The event data from these two triggers cover a time range from $T_{0}+2960$\,s to $T_{0}+4570$\,s. The mask-weighted light curve shows steadily declining emission present when the burst location came into the BAT field of view at $T_{0}+3173$\,s, and extending beyond the available event data range. The time-averaged spectrum from $T_{0}+3302$\,s to $T_{0}+4538$\,s is best fit by a simple power-law model, with $\Gamma = 2.08 \pm 0.03$.  The fluence in the 15--150\,keV band is $(7.4 \pm 0.1) \times 10^{-5}$\,erg\,cm$^{-2}$.

The smooth temporal evolution observed by BAT, together with the large temporal offset between the BAT and GBM triggers, indicate that BAT triggered on the \textit{afterglow} of \event. This marks the first such occurrence of a BAT afterglow trigger in the 18 years of \swift\ operations.

Given the exceptionally bright afterglow, we searched for even later emission in the BAT survey mode data using the \texttt{BatAnalysis} python package (Parsotan et al., in prep.). In individual pointings of survey mode data the afterglow was detected until 2022 October 9 21:55:38~UT ($T_{0}+31$\,ks). We attempted to fit the spectra of each survey dataset with a power-law ({\sc cflux*po}) model in {\sc XSPEC} \citep{XSPEC} to obtain fluxes and photon indices. Non-detections were then analyzed to obtain 5$\sigma$ upper limits following the procedure outlined in \citet{Laha_2022_FRB}.

At later times, the survey data were binned daily and then mosaiced together. The spectrum from each mosaiced image was fitted, and the flux and photon indices derived similar to the procedure above. The results of this analysis are plotted in the top panel of  Figure~\ref{fig:combined_lc}, and listed in Appendix~\ref{sec:BATLC_appendix}. 

\begin{figure*}
    \centering
    \includegraphics[width=\linewidth]{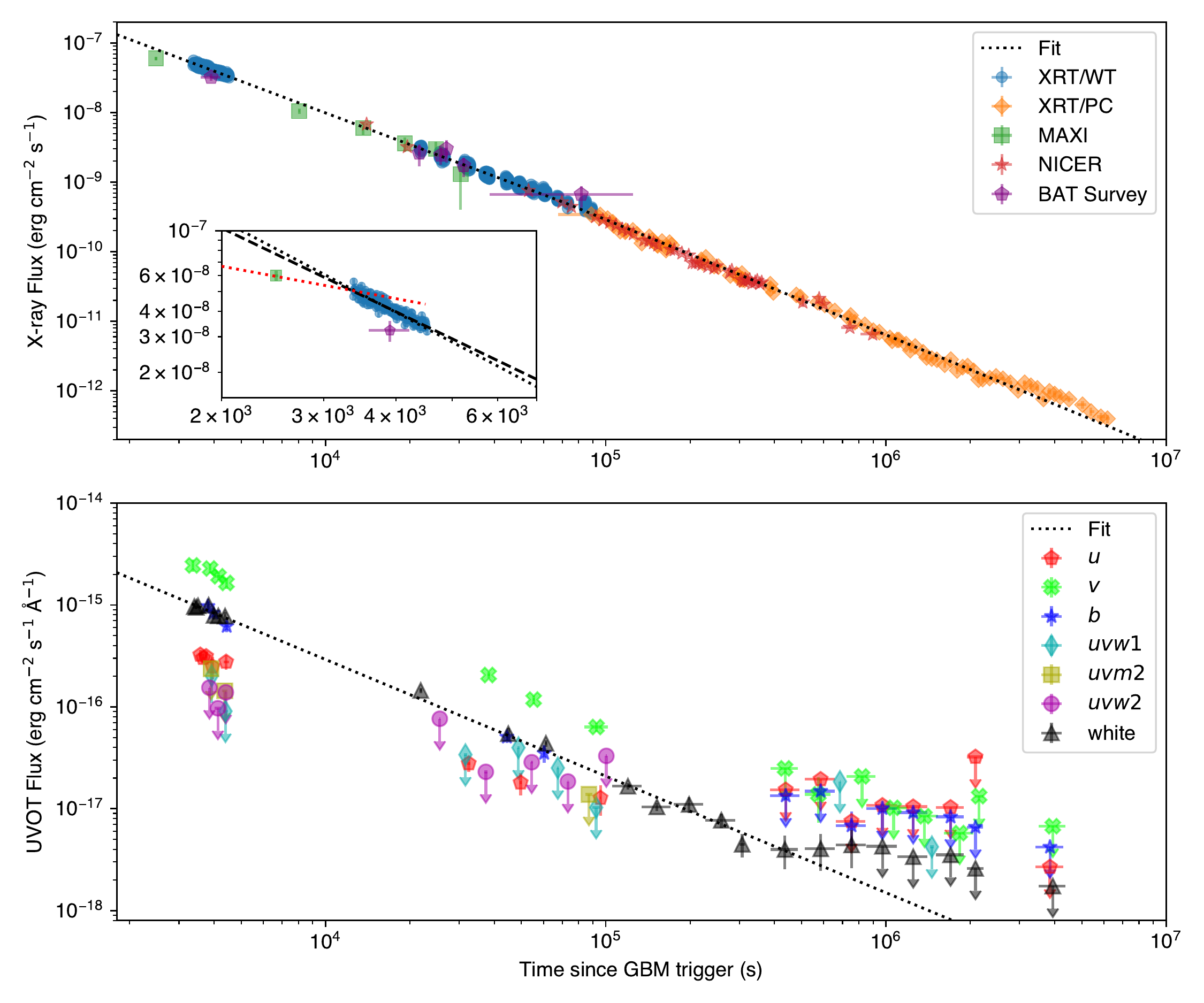}
    \caption{Combined X-ray and UV/optical light curve of \event. The upper panel shows observed X-ray flux  from \maxi{}, \nicer{}, XRT (all 0.3--10.0\,keV) and BAT (14--195\,keV). BAT data are taken from detections of the afterglow in Survey data; note that the final BAT data point combines all observations integrated over 1 day. The dotted line shows a broken power-law fit to the \maxi{}, \nicer{}, and XRT data.    
    The inset figure shows the first \maxi{}, BAT and XRT data, with the black dashed line indicating a fit to the XRT data alone, and red dashed line a fit between the first \maxi{} point and the first XRT detection.
    The lower panel shows 7 filter optical and UV data from UVOT as obtained after the subtraction of late-time template images. The dashed line shows a power-law fit to the white band data. 
        }
    \label{fig:combined_lc}
\end{figure*}

\subsection{\swift\ X-Ray Telescope} \label{sec:xrt}
The XRT began observing \event\ at 14:13:09~UT ($T_0+3370$\,s, 170\,s after the first BAT trigger) and located a bright afterglow in the initial 0.1-s image-mode exposure, after which observations began in Windowed Timing (WT) mode. The initial WT count rate was $910 \pm 40$\,ct\,s$^{-1}$ (all XRT count rates are corrected for the effects of pile-up and hot columns, see e.g., \citealt{Evans07}); making it more than an order of magnitude brighter at this time -- in observed flux -- than any other GRB observed by XRT. Due to the high count rate, the XRT remained initially in WT mode, in which only one-dimensional spatial information is collected. Significant structures are present in the 1-D spatial profiles, with a clear excess compared to the expected point spread function (PSF), which were evolving with time (see Appendix~\ref{sec:xrt_appendix}). This resembles the behavior expected when dust clouds in our Galaxy scatter X-rays from the GRB prompt emission, that were not initially traveling towards Earth, back into our line of sight. 

\begin{figure*}
\centering
\begin{tabular}{lr}
  \includegraphics[width=0.45\linewidth,trim={20 30 55 55},clip]{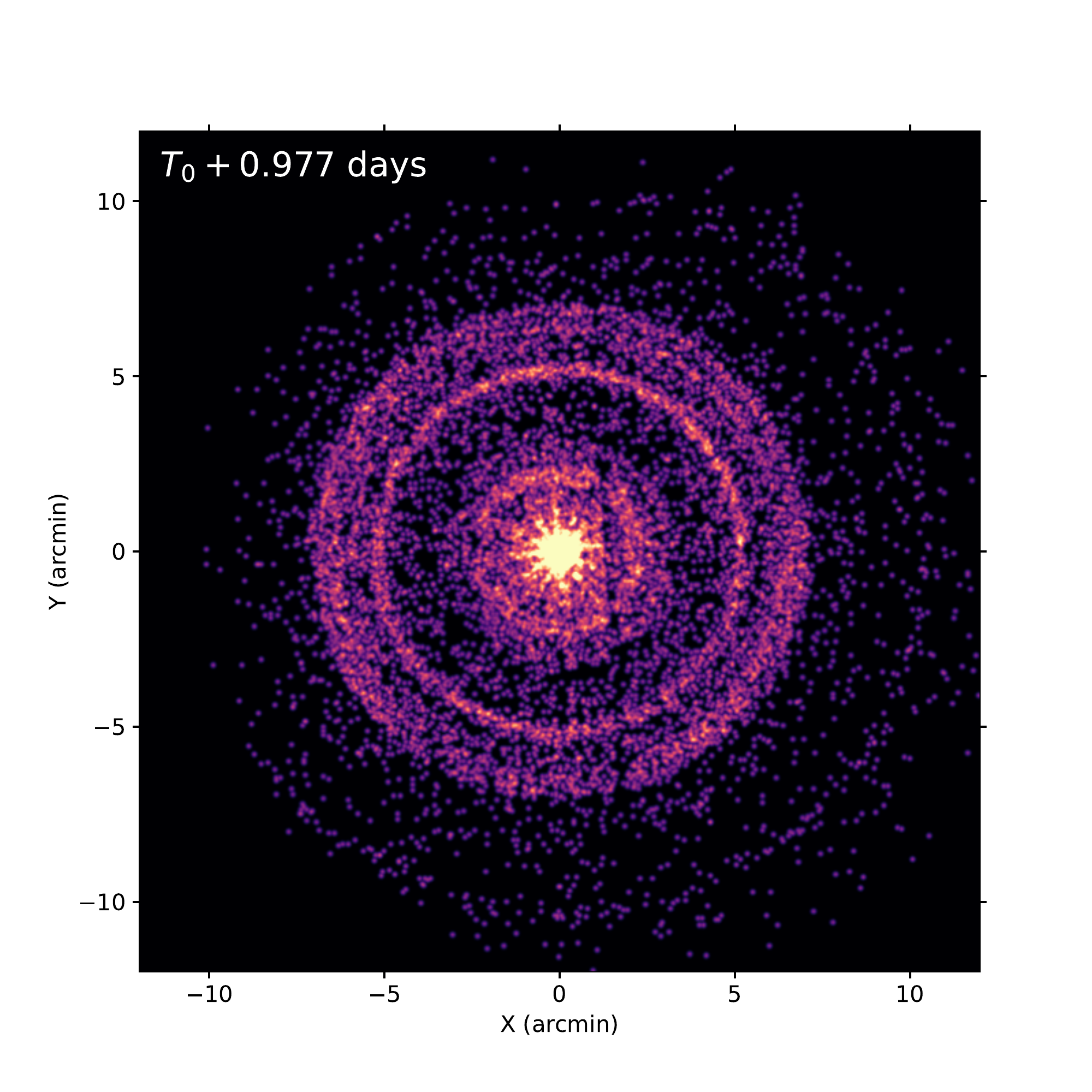} &   \includegraphics[width=0.45\linewidth,trim={20 30 55 55},clip]{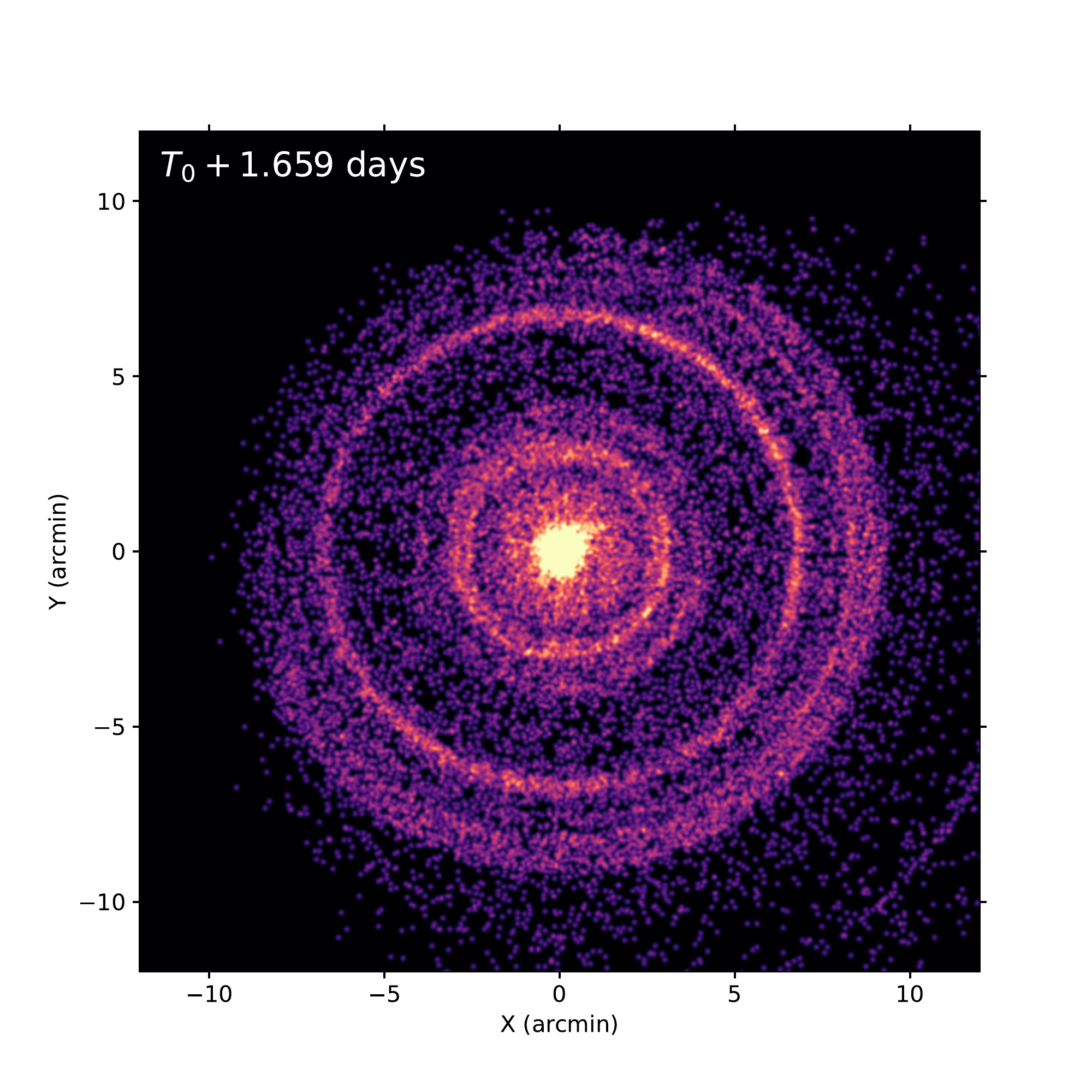} \\
  \includegraphics[width=0.45\linewidth,trim={20 30 55 55},clip]{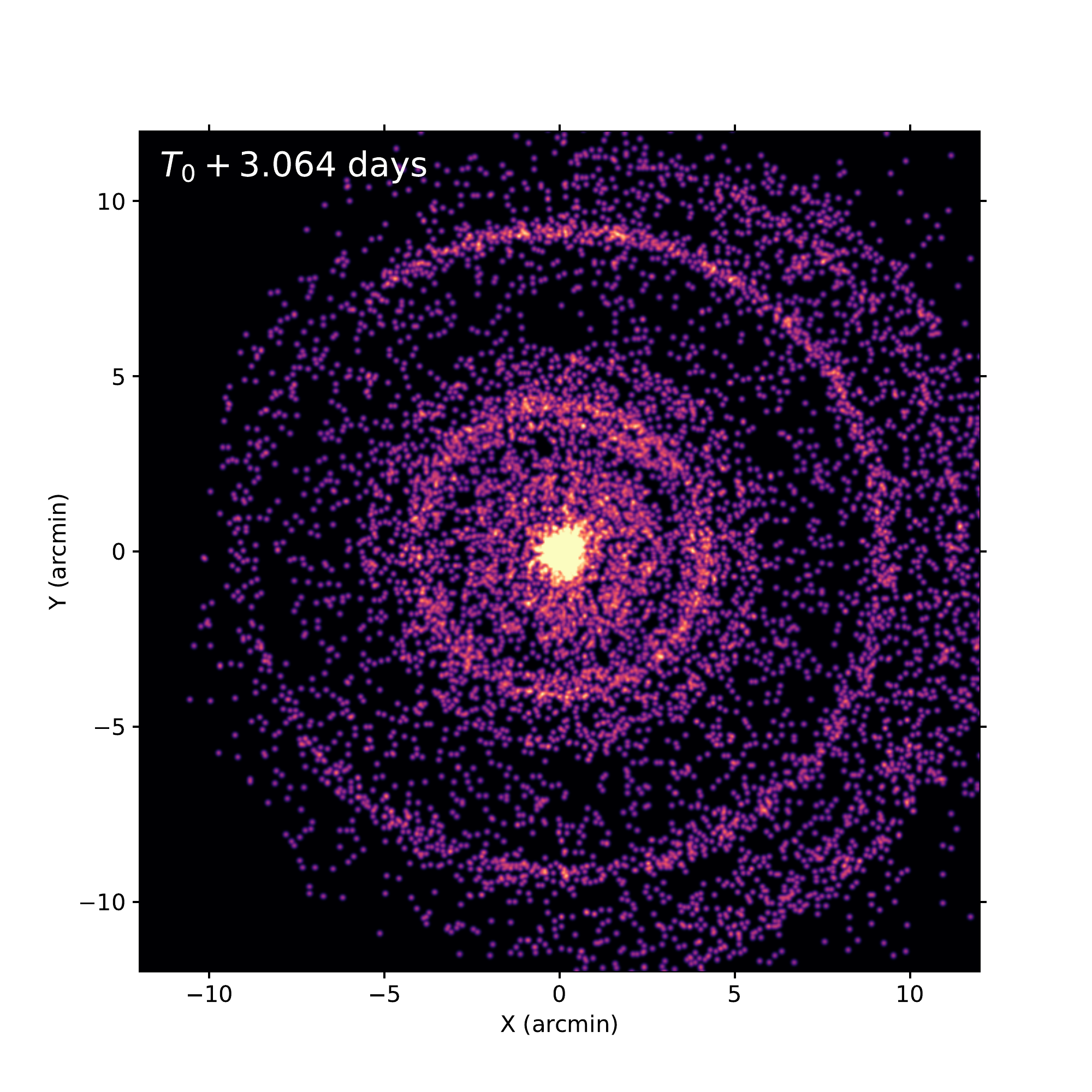} &   \includegraphics[width=0.45\linewidth,trim={20 30 55 55},clip]{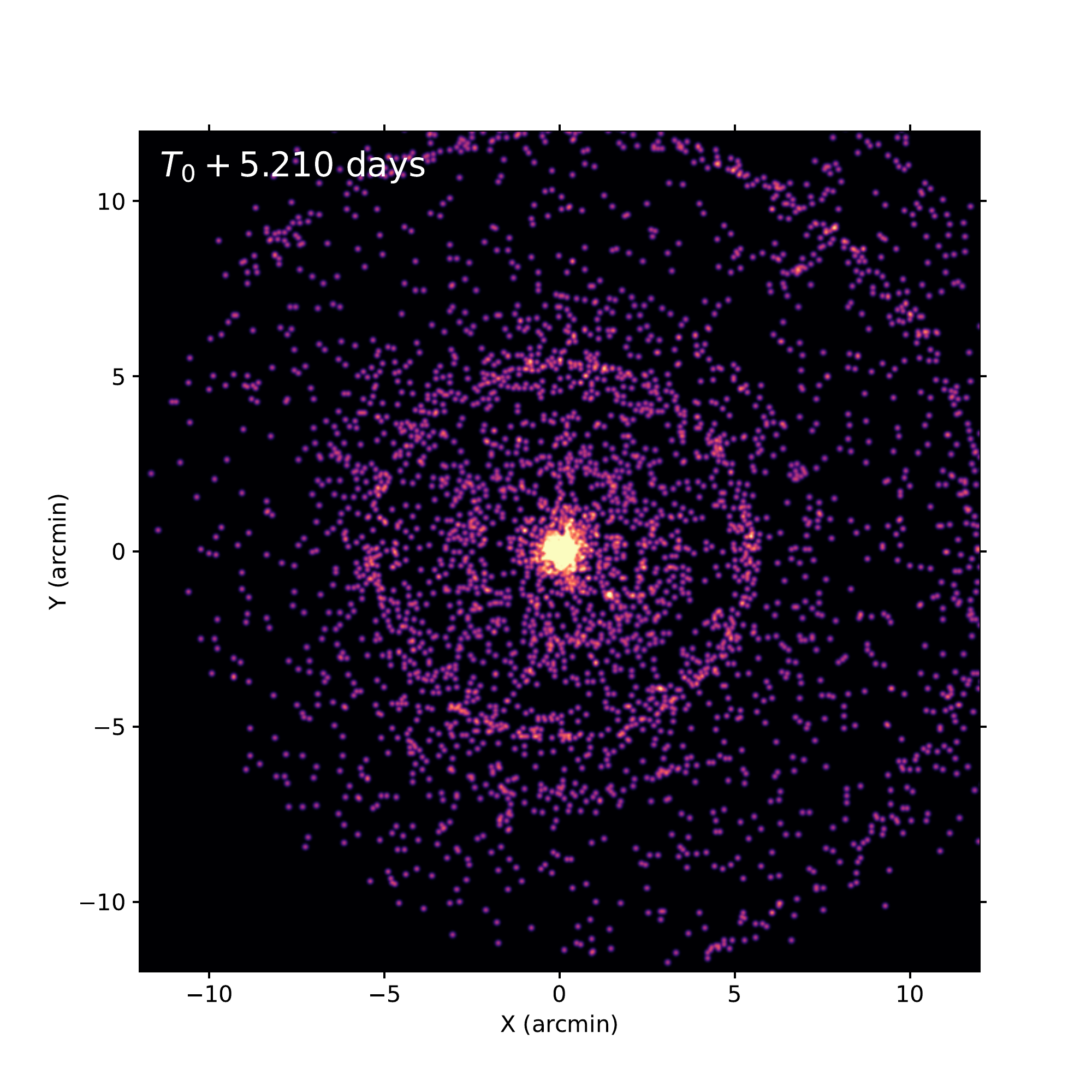}
\end{tabular}
\caption{XRT PC-mode $0.8-5.0\,\mathrm{keV}$ images from ObsIDs 01126853004 (top left),  01126853006 (top right), 01126853008/09 (bottom left) and 01126853010/11 (bottom right), illustrating the echo expansion with time.
The echo event radial positions were scaled by $t^{-0.5}$ to each observation mid-time (given in the upper left corner of each image in days since GBM trigger) to counter the halo expansion within each observation.}

\label{fig:xrtrings}
\end{figure*}

At $T_{0} + 89$\,ks, the GRB had faded sufficiently that the XRT automatically switched to Photon Counting (PC) mode. The 2-D image from this observation, shown in Figure~\ref{fig:xrtrings}, confirmed the presence of a complex series of expanding, bright rings associated with a dust-scattering echo \citep[see also][]{ATel15661}, the properties of which are discussed in \S~\ref{sec:dust}.

The presence of scattered emission complicates the data analysis, as it can contribute events to the regions over which source and background counts are accumulated. Both the intensity and spectrum of the rings are spatially variable, thus the selected `background' region may in fact not be representative of the background within the source region. As a result, the automated XRT analysis \citep{Evans07,eva09} needs some modification. For PC mode this is relatively simple, as we have full 2-D imaging and the innermost rings were reasonably well separated from the GRB itself. We restricted the source region to a radius of 20 pixels (47\arcsec), and manually defined a background region as free from dust contamination as possible (identified from a stacked image of all PC mode data). 

For WT mode this is more problematic since the CCD is read out in columns, rather than pixels. After investigating the variation in the echo contribution to WT-mode data, we modified the default WT extraction regions to minimize its impact (see Appendix~\ref{sec:xrt_appendix}). We found that the WT flux is accurate to $\sim$6\%. 
We further verified this by extracting a light curve using only data above 4\,keV, where dust scattering becomes less efficient, given the roughly $\nu^{-2}$ dependence of the scattering cross section; only minimal changes in the light curve shape are observed. On the other hand, the dust has more significant impact on the WT spectra (Appendix~\ref{sec:xrt_appendix}), resulting in an increase in the best-fit photon index up to $\sim$7\% and a significant increase to the best-fit absorption column (up to $\sim$27\%). 

We modified the settings for the automated light curve\footnote{\url{https://www.swift.ac.uk/xrt\_curves/01126853/}.} to limit the source extraction region to 20 pixels radius, and to use the PC-mode background region (described in the Appendix~\ref{sec:xrt_appendix}); due to the unusual brightness of the source we also increased the number of counts per bin by a factor of ten compared to the default (see \citealt{Evans07} for details). We fit the time-averaged PC mode spectrum using {\sc XSPEC} with a power-law model and two absorption components. The first of these was a {\sc tbabs} fixed to the Galactic value of 5.38\tim{21} \cms\ \citep{wil13}; the second was a {\sc ztbabs} with \nh{} free and redshift fixed at 0.151.
The fit yields $\nh=(1.4\pm0.4)\tim{22}$ \cms, with a photon index $\Gamma=1.8\pm0.2$. We used these results to convert the light curve into observed 0.3--10\,keV flux. The resulting flux measurements are shown in Figure~\ref{fig:combined_lc}.

To investigate possible spectral evolution, we extracted a series of time-resolved spectra using the `Add time-sliced spectrum' option on the UKSSDC website\footnote{\url{https://www.swift.ac.uk/xrt\_spectra/01126853/}.} \citep{eva09}, which ensures that the modifications to the default processing parameters, described above, are employed. From the WT-mode data we created one spectrum from the first spacecraft orbit ($T_0$+3.4--4.5\,ks), and then spectra in 10\,ks chunks from $T_0$+10--50\,ks. For PC mode, we extracted spectra over the intervals $T_0$+67--200\,ks, 200--400\,ks, 400--1000~ks and $>1000$~ks. For WT mode we used only grade 0 events, as is recommended for very absorbed objects\footnote{\url{https://www.swift.ac.uk/analysis/xrt/digest_cal.php\#abs}}. We fitted these spectra independently with the model defined above; the results are shown in Table~\ref{tab:spectra}.
The photon index ranges between $\sim$~1.6 and $\sim$~1.9 with the harder profiles observed during the first block of observations (from 3.4 to 4.5\,ks) and around 1-2 days (from 68 to 175\,ks) after $T_0$.

\begin{deluxetable*}{ccccccccc}
\tablecaption{Spectral parameters determined from fitting a power-law model to the time-resolved X-ray data from \swift{}, \maxi{} and \nicer{}.\label{tab:spectra}}
\tablehead{	 \colhead{$T - T_0$}	&	\colhead{Exposure}		&	\colhead{Intrinsic \nh{}}	& \colhead{Photon} & \colhead{Fit stat. /} & \colhead{Observatory /}\\
(ks)& (ks)	&  (10$^{22}$cm$^{-2}$) &  index &  d.o.f & Detector}
\startdata
 2.5  & 0.047	& --- &$1.75 \pm 0.09$		 & 307/322 & \maxi{}/GSC\\
 $3.4 - 4.5$                    & 1.2           &   $1.28\pm0.04$                &  $1.61\pm0.02$ & 1249 / 959  & XRT/WT \\
 $8.0$	& 0.047 & --- &$2.07^{+0.28}_{-0.26}$ 	 & 145/166 & \maxi{}/GSC\\
$10.0 -20.0$                     & 0.5           &   $1.39\pm0.09$                &  $1.89\pm0.05$ & 859 / 840 & XRT/WT\\
$14.0$   & 0.029 & --- & $1.89\pm0.16$  & 26.9/35 & \nicer{}/XTI\\
$19.5$   &0.062 & --- &$1.84\pm0.15$ & 23.6/36 &\nicer{}/XTI\\
$20.0 - 30.0$                      & 2.9           &   $1.36\pm0.05$                &  $1.84^{+0.02}_{-0.04}$ & 995 / 922 & XRT/WT\\
$30.0 - 40.0$                      & 1.7           &   $1.25\pm0.07$                &  $1.85\pm0.04$ & 800 / 840 & XRT/WT\\
$40.0 - 50.0$                      & 2.8           &   $1.20\pm0.06$                &  $1.84\pm0.03$ & 875 / 873 & XRT/WT\\
$52.6 - 98.0$   & 2.1 & ---& $1.85\pm0.07$ & 131 / 153 & \nicer{}/XTI	\\$68-175$                     & 17            & $1.27^{+0.16}_{-0.15}$         & $1.65\pm0.08$    & 655 / 727  & XRT/PC   \\
$103.0 - 198.0$ & 6.5	& --- & $1.79\pm+0.06$ & 144 / 168 & \nicer{}/XTI	\\
$203.3 - 360.5$	& 9.8 & --- &1.61$^{+0.10}_{-0.09}$  & 94 / 94 & \nicer{}/XTI	\\
$216-394$                    & 12            & $1.26^{+0.18}_{-0.17}$         & $1.80\pm0.10$    & 632 / 665 & XRT/PC   \\
$404-954$                    & 26            & $1.20\pm0.18$                  & $1.92\pm0.10$    &  501 / 622  & XRT/PC   \\
$505.0 - 961.6$ & 2.7	& --- & $1.82\pm0.19$ & 58.1/67& \nicer{}/XTI \\
$1005-6285$   & 0.24           & $1.00\pm0.15$                  & $1.91\pm0.09$    & 723 / 668  & XRT/PC      \\
\enddata
\tablecomments{\nicer{}/XTI and \maxi{}/GSC data are fit in the energy range $4-10$\,keV and $4-20$\,keV, respectively, to avoid contamination from unresolved dust echo contamination, with intrinsic \nh{} fixed to 1.29\tim{22}~cm$^{-2}$. XRT spectra are fit in the range 0.3 -- 10\,keV, allowing for measurement of intrinsic \nh{}. Note that fit statistics are C-stat \citep{Cash79} for \maxi{} and \swift{} data, \nicer{}{} are $\chi^2$. Galactic \nh{} is fixed to a value of 5.38\tim{21} \cms\ \citep{wil13}.}
\end{deluxetable*}

\subsection{\swift\ Ultraviolet/Optical Telescope} \label{sec:uvot}
The UVOT began settled observations of the field of \event\ at 2022 October 9 14:13:17~UT, $T_{0}+3.4$\,ks (179\,s after the first BAT trigger; \citealt{GCN:32632,GCN:32656}).
A nearby star, 5\arcsec\ away from the GRB, contaminates the photometry when using the standard circular aperture of 5\arcsec\ radius. In order to minimize the star's contribution to the measurement of the afterglow, we use a 2.5\arcsec\ aperture to extract source counts. To be consistent with the UVOT calibration, these count rates were then corrected to 5\arcsec\ using the curve of growth contained in the calibration files. Background counts were extracted using two circular regions of radius 15\arcsec\ located in source-free regions. The count rates were obtained from the image lists using the \swift{} tool \texttt{uvotsource}.

As \event\ fades, even using a small aperture, the contribution from the nearby star increases in dominance. In order to estimate the level of contamination, for each filter we combined the late time exposures between $T_{0}+(3.4 \times 10^{6})$\,s and $T_{0} +(4.4 \times 10^{6})$\,s. We extracted the count rate in the late combined exposures using the same 2.5\arcsec\ aperture and applied an aperture correction to 5\arcsec. These 
were subtracted from the source count rates to obtain the afterglow count rates. The afterglow count 
rates were converted to magnitudes using the UVOT photometric zero points \citep{poole,bre11}, as well as to fluxes. To improve the signal-to-noise ratio, the count rates in each filter were binned using $\Delta t/t$ =0.2. The final photometry is plotted in the bottom panel of Figure~\ref{fig:combined_lc} and listed in Appendix~\ref{sec:UVOT_appendix}.

\subsection{\maxi{}} \label{sec:maxi}
\maxi{}, which is onboard the International Space Station (ISS), performs scanning observations of about 80\% of the sky every 92~min. 
Prior to the GBM detection of the prompt emission from \event{}, the two Gas Slit Camera (GSC; \citealt{mihara11}) units,  GSC\_4 and GSC\_5, scanned the source region at 12:25~UT on 2022 October 9 (\tz$-3.1$\,ks), 
and no enhancement was seen at that time.
The 2--10\,keV 3$\sigma$ upper limit is 0.074 ct cm$^{-2}$ s$^{-1}$, corresponding to a flux of approximately 7.9\tim{-10} \ecs{}. 
At 13:58~UT (\tz$+2.5$\,ks) the GSCs detected a very bright transient with a 2--10\,keV flux of $5.97\pm0.19$ ct cm$^{-2}$ s$^{-1}$ ($\sim6.5$\tim{-8} \ecs{}; \citealt{ngrgrb}).
Unfortunately, at the time the data were collected, there was a loss of the signal to the ground, and the data were 
stored in the High Rate Communication Outage Recorder in the ISS, leading to a delay in reporting the transient.
Before downloading these stored data at around 15:55~UT, the \maxi{}/GSC Nova-Alert System \citep{negoro16} triggered on the source at 15:31:03~UT (\tz$+8.0$\,ks), about 81 min after the BAT trigger. 
In this second observation by \maxi{}, the 2--10\,keV flux had dropped to $1.00\pm0.08$~ct~cm$^{-2}$~s$^{-1}$ (about 1\tim{-8} \ecs{}).

Thanks to the exceptional brightness, the GSCs significantly detected the afterglow during four scans that followed the second detection, to 21:41~UT, six scans in total (e.g., \citealt{kobagcn}). Typically, GRBs are not detected beyond the first one or two scans, taken at $\sim92$ min intervals .
After the six detections, the sky region containing the source was not observable by \maxi{} until 15:31~UT on 2022 October 11.

We performed a spectral analysis using data obtained from GSC\_4 and GSC\_5 (for details see Appendix~\ref{sec:MAXIappendix}). We first attempted a standard analysis using the same model applied to the XRT spectra, i.e., a power-law with two absorption components, with a fixed Galactic absorption column density of $\nh{} = 5.38\tim{21}$~cm$^{-2}$. However, we obtained steeper photon indices and/or lower intrinsic absorption column densities
than those of the XRT (see Table~\ref{tab:gscspec}). 
If we fixed $N_\mathrm {H,zabs}$ at the weighted mean value obtained with the XRT at \tzks{+(3.4-50)}, 1.29\tim{22} cm$^{-2}$,
even steeper photon indices were obtained.

These discrepancies can be explained by the GSC spectra containing dust scattered soft X-rays, which can not be spatially resolved by the GSC. Mitigation of the dust scattering process is discussed in Appendix~\ref{sec:MAXIappendix}; to summarize, an additional spectral component to account for the scattered emission was included with a power-law index derived from XRT PC-mode data and the associated flux allowed to vary as a free parameter. The afterglow photon index was fit as a free parameter in the data from the first two scans, but fixed at the weighted mean XRT value at \tzks{+(10-50)} due to poor statistics. The resulting spectral parameters are displayed in Table~\ref{tab:spectra}, and flux measurements (extrapolated to the 0.3--10.0\,keV band) are plotted in Figure~\ref{fig:combined_lc}.

\subsection{\nicer{}} \label{sec:nicer}
\nicer{} received notification of this GRB from \maxi{} on the ISS via OHMAN (On-orbit Hookup of \maxi{} and \nicer{}) at 14:10:57~UT, but could not begin regular observations due to poor visibility until $T_0+52.870$\,ks. Only two observations were possible during this early phase. The first and second observations were at $T_0+14.003$\,ks and $T_0+19.560$\,ks with an exposure time of 29 s and 62 s, respectively. After $T_0+52.870$\,ks, \nicer{} intermittently made observations over a period of about 42 days.

\nicer{}’s X-ray Timing Instrument (XTI) consists of 56 modules, each comprising an X-ray concentrator optic and silcon drift detector \citep{NICER}.  During these observations 52 XTI modules were operational. We reduced the data with the {\tt nicerl2} command and generated level 2 cleaned events. To avoid potential contamination of instrumental noise, we further excluded XTI module numbers 14 and 34. We defined a time interval in which an exposure time of more than 200 s can be continuously secured as a one-time interval and created an energy spectrum at each time interval (except for the first two observations) with good statistics. 

To select a time period with less background influence due to charged particles, we also checked the distribution of the rate of 12--15\,keV photons, where the optics have less effective area. Most data are distributed around 0.048 ct s$^{-1}$ and can be represented by a normal distribution with a standard deviation of 0.018 ct s$^{-1}$. Therefore, we removed data with a rate greater than 2$\sigma$ from the mean in the 12--15\,keV range after $T_0$+100\,ks seconds where the background contribution becomes large
($> 0.084$ ct s$^{-1}$). We also did not use data after $T_0$+1000\,ks because the background rate increased due to a geomagnetic storm. The background spectral model is estimated by the 3C50 background model \citep{remillard22} using the {\tt nibackgen3c50} command.

To explore the spectral evolution as in \S\ref{sec:xrt}, we divided the \nicer{} data into six time intervals and fitted the time-resolved spectra with a power-law model and the same two absorption components. Since \nicer{} is a non-imaging detector, it cannot remove photons originating from the dust echo within its 5\arcmin\ diameter field-of-view. To evaluate the impact of the dust echo, we fitted the data in two energy ranges, 1--8\,keV and 4--8\,keV bands. 
Since the data fitting in the 4--8\,keV range does not allow us to constrain the intrinsic \nh{} value, we fixed it to 1.29$\times 10^{22} $\,cm$^{-2}$ which is the weighted mean value obtained by XRT at $T_0$+(10.0--961.6)\,ks (Table~\ref{tab:spectra}). Comparing the fitting results in the two energy bands, the results for the 4--8\,keV band show that the photon indices are systematically harder than those for the 1--8\,keV band results and consistent with the XRT results. Therefore, we conclude that 4--8\,keV light curve is much less affected by the dust echo, and report these results in Table~\ref{tab:spectra}. 

\section{Analysis} \label{sec:analysis}

\subsection{Dust Scattering Echo} \label{sec:dust}

While complicating the point source analysis of the afterglow emission, the dust scattering echo contains a rich amount of information about the intervening Galactic dust along the line of sight to \event{}. Dust echoes from GRBs offer the unique opportunity to measure distances to dust clouds geometrically, with a precision limited only by the angular resolution of the telescope. This is because the distance $D_{\rm GRB}$ to the X-ray source is effectively infinite, relative to the distance to the dust. This allows us to eliminate the source distance from the scattering angle equation \citep[e.g.,][]{heinz15} and solve it directly for the distance $D_{\rm dust}$ to the dust (using the small angle approximation):
\begin{equation}
    D_{\rm dust} = \frac{D_{\rm GRB}}{1 + \frac{D_{\rm GRB}\theta^2}{2c\Delta t}}\approx \frac{2c\Delta t}{\theta^2}\label{eq:dustdistance}
\end{equation}  
where $\Delta t$ is the time delay between the GRB and the exposure and $\theta$ is the off-axis angle of the ring relative to the position of the GRB.

Because the vast majority of the fluence of the GRB was concentrated in the prompt GRB emission, to lowest order, we can approximate the input light curve that caused the echo as a delta function in time. This implies that there is a single, unique $\Delta t$ for each photon detected by the XRT, and thus, each photon can be referenced to a specific dust distance in eq.~\ref{eq:dustdistance}. This makes the analysis of this echo substantially simpler than in comparable echoes from Galactic X-ray transients.

Details of our analysis of the dust echoes from \event{} are provided in Appendix~\ref{sec:dustappendix}. Figure~\ref{fig:columndensityhistogram} shows the best fit column density histogram for a standard Mathis, Rump, \& Nordsieck (MRN) model \citep{mathis77}, assuming a soft X-ray fluence of ${\mathcal F}_{\rm 0.8-5\,keV} = 2.1\times 10^{-3}\,{\rm erg\,cm^{-2}}$ (Lesage et al.\ in prep.). The figure shows clear evidence for several dust components on the near side of the Galaxy, with the largest column densities found within distances of about 1 kpc, as expected given the Galactic latitude of \event{}.

\begin{figure}
    \centering
    \includegraphics[width=\linewidth]{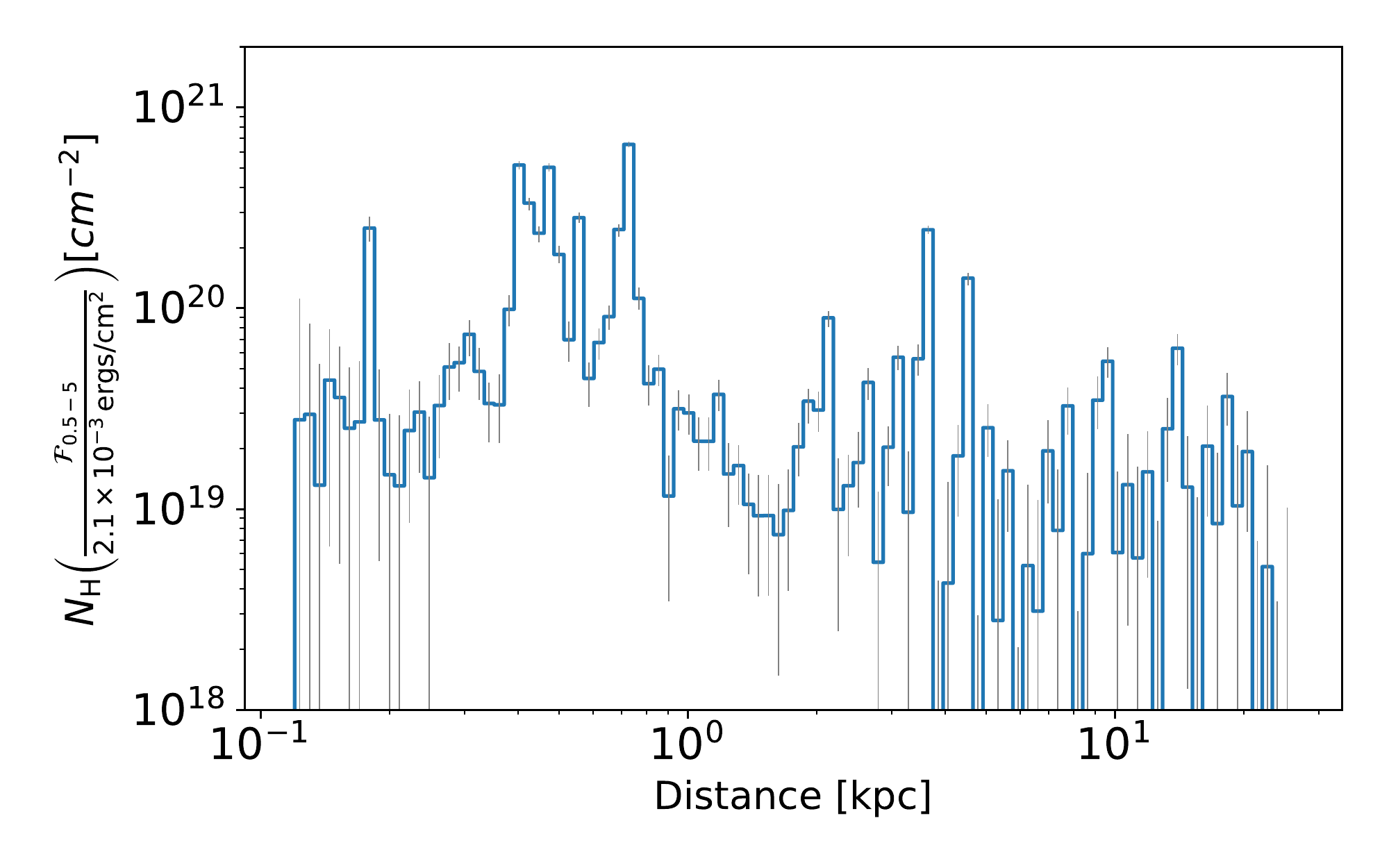}
    \caption{Dust column density (expressed as equivalent hydrogen column density, assuming solar abundances) determined from fitting dust scattering models to the radial intensity profiles as a function of distance, for a standard MRN model. The dependence on the uncertain soft X-ray fluence of \event{} is shown in the y-axis label. Vertical gray lines indicate 1$\sigma$ error bars.}
    \label{fig:columndensityhistogram}
\end{figure}

However, there is also clear evidence for dust at larger distances, located well above the Galactic disk (see also \citealt{Negro+2023}). While the column densities of these dust concentrations are small, the large cross section at the correspondingly small scattering angles of this more distant dust makes the scattering from these clouds detectable in the radial intensity profiles, as can be seen in Figure~\ref{fig:pchaloradialprofiles}, for example. As such, the dust echo from \event{} proves that echo tomography is exquisitely sensitive to small dust concentrations at large distances that are not typically measurable with other techniques.

The largest column density cloud is located approximately between 400 and 600 pc distance, with a very nearby dust cloud at a distance of about 200 pc and a third cloud at distance approximately 700 pc. Naturally, these structure are very local compared to the large scale Galactic structure. The very rapid detection of the echo by the XRT allows us to map very nearby dust typically hard to study using dust echoes.

We find a total Galactic column density of $N_{\rm H} \sim 5.9\times 10^{21}\,{\rm cm^{-2}}$, broadly consistent with the Galactic column density of $5.38\times 10^{21}\,{\rm cm^{-2}}$ found by \citet{wil13}. Given that the largest uncertainty in this value derives from the poorly constrained estimate of the soft X-ray fluence of \event{}, we can infer that our assumed fluence of ${\mathcal F}_{\rm 0.8-5\,keV} = 2.1\times 10^{-3}\,{\rm ergs\,cm^{-2}}$ is likely correct within roughly $10\%$.

We find that a simple MRN distribution provides an excellent fit to the temporal evolution of the radial intensity profile, without the need for more complex dust chemistry (such as ice mantles). In this sense, the dust towards \event{} appears to be typical for interstellar dust seen along sightlines to other recent X-ray light echoes, such as the echo from V404 Cyg \citep{beardmore16,heinz16,vasilopoulos:15}.

\subsection{Broadband Spectral Energy Distribution} \label{sec:SEDs}
The combined XRT+\maxi{}+\nicer{} dataset allows us to explore the soft X-ray afterglow evolution in great detail. In the earliest soft X-ray spectra ($T_{0}+2.5$--10\,ks), the afterglow exhibits strong spectral evolution, with the XRT WT-mode power-law photon index increasing (i.e., getting softer) by $\sim17$\% (Table~\ref{tab:spectra}). This change is much larger than the inferred WT-mode systematic uncertainty (Appendix~\ref{sec:xrt_appendix}), and is corroborated by the 2--20\,keV \maxi{} spectra derived around this time (Appendix~\ref{sec:MAXIappendix}).  

From $T_{0}+10$--50\,ks, the spectra show no definitive signs of evolution; there are hints at a reduction in the host galaxy absorption throughout the XRT WT-mode data, although this trend is reversed when XRT switched to PC mode. Given the differences between the WT- and PC-mode results are much less than the dust-induced systematics on the WT results (Appendix~\ref{sec:xrt_appendix}), and no similar spectral evolution is observed in the \nicer{} spectra, we do not believe there is supportable evidence for spectral evolution at these times. 

On the longer timescales probed by the XRT PC-mode data, clear spectral evolution is seen and takes the form of an increasing (i.e., softer) photon index, which is confirmed by the 4--8\,keV \nicer{} spectra (Table~\ref{tab:spectra}).

To constrain the broadband spectral behavior, we
built and fit two spectral energy distributions (SEDs) at $T_{0}+4.2$\,ks and $T_{0}+43$\,ks, following the procedure outlined in \cite{sch10,schady07}, using data from the BAT survey mode, XRT and UVOT. 
For the 4.2\,ks SED we use data in the range 2.5--4.7\,ks. For the 43\,ks SED we used data in the range 23--103\,ks. Details of the SED construction are provided in Appendix~\ref{sec:SEDappendix}.

The SEDs were fit using {\textsc XSPEC}. We tested two different models for the continuum: a single power-law, and a broken power-law with the change in spectral slope fixed to be $\Delta \beta = 0.5$ (corresponding to the expected change in spectral slope caused by the synchrotron cooling frequency; \citealt{sar98}). In each of these models, we also included two dust and gas multiplicative components to account for the Milky Way and host galaxy dust extinction and photoelectric absorption ({\sc tbabs}, {\sc ztbabs}, {\sc zdust})\footnote{We used {\sc zdust} for both the Galaxy and host galaxy dust components, but with the redshift set to zero for the Milky Way dust component.}. We also include a component for attenuation by the intergalactic medium ({\sc zigm}). The Galactic components were frozen to the reddening and column density values from \cite{schlegel} and \cite{wil13}, respectively. In the analysis, we found consistent results between Milky Way, Small Magellanic Cloud (SMC) and Large Magellanic Cloud dust extinction laws, therefore we only report the fits using the SMC dust extinction law. 

\begin{figure*}
    \centering
    \includegraphics[width=0.45\linewidth,trim=3cm 8cm 4cm 8cm,clip]{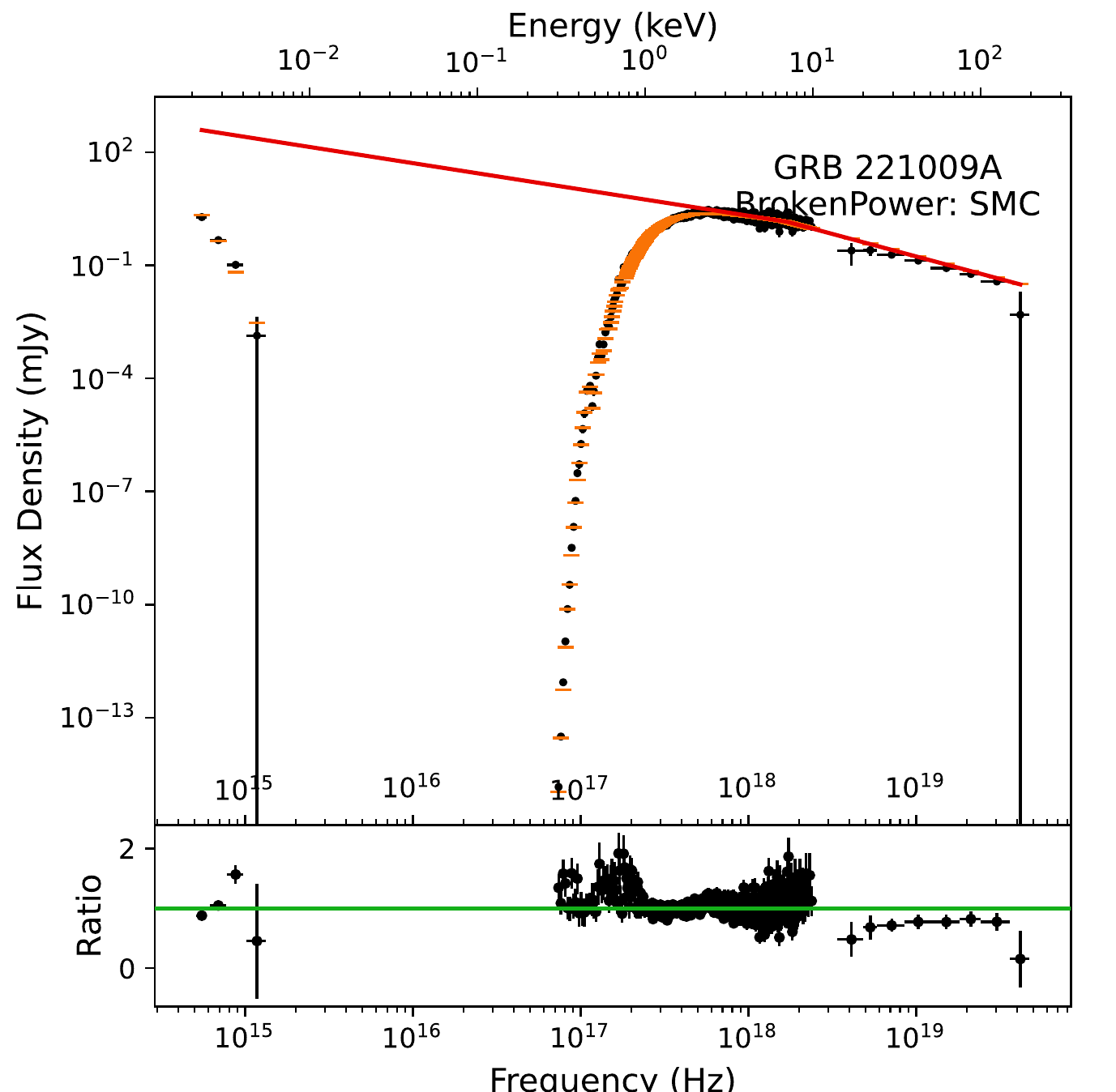}
    \includegraphics[width=0.45\linewidth,trim=3cm 8cm 4cm 8cm,clip]{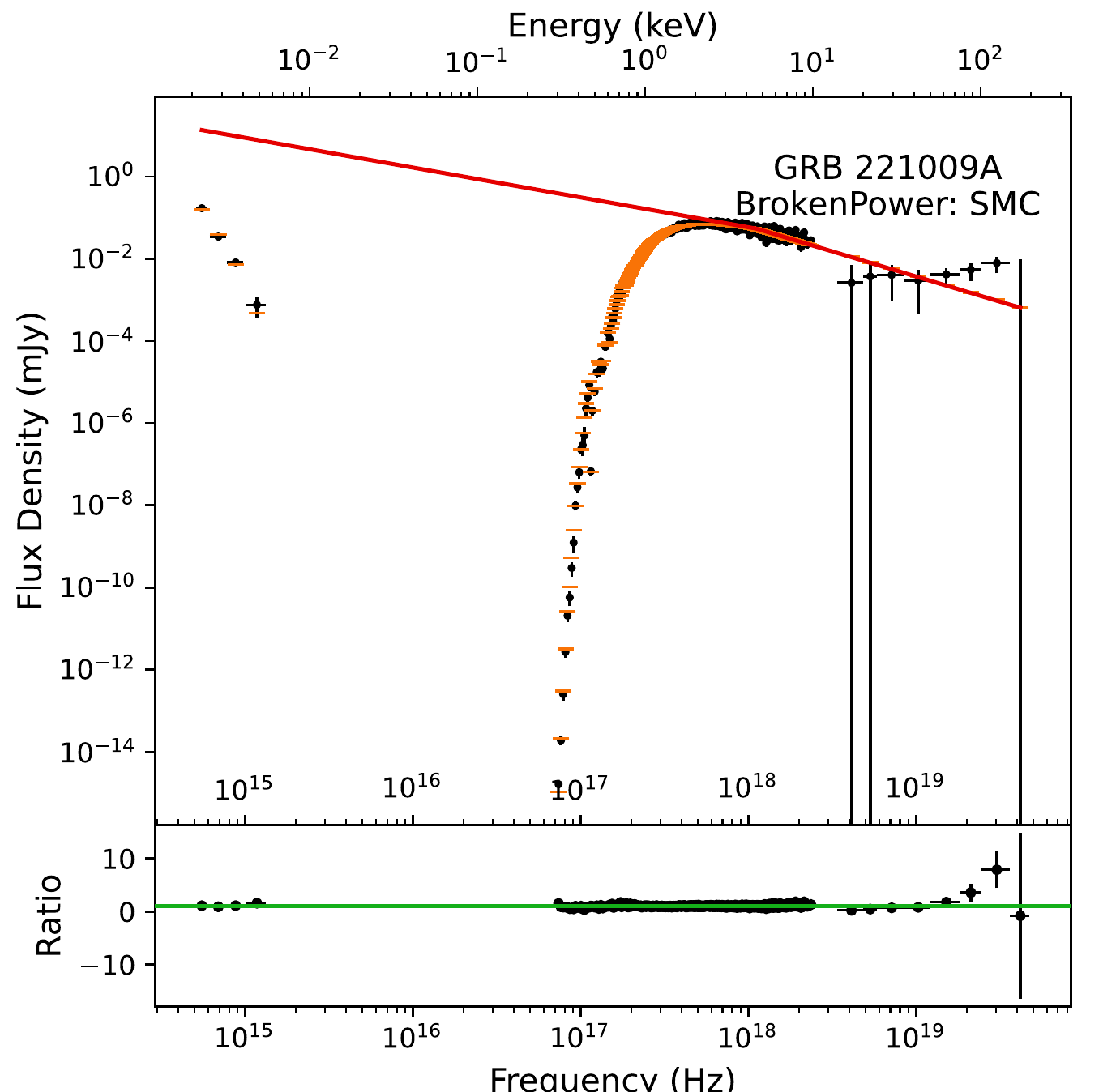}
    \caption{Spectral energy distributions at $T_{0}+4.2$\,ks and $T_{0}+43$\,ks, containing BAT, XRT and UVOT data. The top panel displays the best fit unabsorbed broken power-law model (red) together with the data (black) and the data folded with the model (orange). The bottom panel shows the ratio between the data and the model folded through the instrument response.
}
    \label{fig:seds}
\end{figure*}

Due to the uncertain contribution to the XRT WT spectra by the dust rings, we extracted WT spectra using the 5-20 pixel annular extraction region identified in Appendix~\ref{sec:xrt_appendix} for the spectrum for the early SED, and a 10 pixel radius circular extraction region for the later time SED, chosen to minimize the fraction of dust scattered emission in the extraction region, with a background spectrum taken from a 75–125 pixel annular region.\footnote{See https://www.swift.ac.uk/analysis/xrt/backscal.php.} Since the intrinsic \nh{} in the WT fits is sensitive to the dust contamination (see Appendix~\ref{sec:xrt_appendix} and Table~\ref{tab:xrtCheckSpec}), we also provide fits using BAT and UVOT alone. The results of the analysis are given in Table~\ref{tab:sedfits} and plotted in Figure~\ref{fig:seds}.

Examining the SED at $T_{0}+4.2$\,ks including BAT, XRT and UVOT data, we find that a broken power-law is preferred, with the $F$-test finding a statistical improvement of $\gg 5\sigma$. The break energy required is at $\sim7$\,keV. At the same epoch, excluding the XRT data we find again a broken power-law model is preferred, with the $F$-test suggesting the improvement is $> 3\sigma$. The photon index for the broken power-law model is consistent with the fits including the XRT spectral files. However, the intrinsic dust extinction is lower: $0.23 \pm 0.05$\,mag compared to $0.51 \pm 0.03$\,mag including all data. Not surprisingly, excluding the XRT data greatly increases the uncertainty in the break frequency.

For the SED at $T_{0}+43$\,ks, a broken power-law is again preferred, with the $F$-test suggesting a break is statistically required at $>3\sigma$. The break frequency remains in the X-ray bandpass, with evidence for a decline in energy over this period. The fits still prefer a significant host galaxy reddening, but the exact value of $E(B-V)_{\mathrm{host}}$ again varies depending on the inclusion or exclusion of the XRT data.

To summarize, both at $T_{0}+4.2$\,ks and $T_{0}+43$\,ks, we find strong evidence favoring a broken power-law model, with the lower energy photon index $\Gamma \approx 1.7$. Both epochs favor a break energy in the X-ray bandpass, and a modest amount of intrinsic absorption/reddening in the host galaxy: $E(B-V)_{\mathrm{host}} \approx 0.3$--0.5\,mag; $N_{H} \approx (1.1$--1.4)$\times 10^{22}$\,cm$^{-2}$. More precise constraints, however, are precluded due to the complications resulting from the dust scattering echoes.

\begin{deluxetable*}{cccccccc}
\tablecaption{Fits to the two spectral energy distributions, at $T_{0}+4.2$\,ks and $T_{0}+43$\,ks, built using UV/optical (U), X-ray (X) and gamma-ray (B) data. Both epochs were fit with a power-law (POW) and a broken power-law (BKPOW) continuum, accounting for Galactic and host galaxy gas and dust. The columns are time of the SED; model; host galaxy equivalent column density N$_{H,X,\mathrm{host}}$; host galaxy reddening, E(B-V)$_{\mathrm{host}}$; power-law photon index or the first photon index of the broken power-law model $\Gamma$, ($\Gamma_2$ is fixed to be $\Gamma+0.5$); break energy for the broken power-law spectral models E$_{bk}$, $\chi^2$ \& degrees of freedom (d.o.f); and the null hypothesis probability.\label{tab:sedfits}}
\tablehead{SED & Model & N$_{H,X,\mathrm{host}}$ & E(B-V)$_{\mathrm{host}}$ & $\Gamma$ & E$_{bk}$ & $\chi^2$ (d.o.f) & Null Hypothesis\\ 
& & ($10^{22}$~cm$^{-2}$) & (mag) & & (keV) & & Probability}
\startdata
B+U 4.2\,ks & POW    & --& $0.26\pm0.05$ & $1.74\pm0.02$ & --  &  16 (9)  & 7.3e-02  \\
B+U 4.2\,ks & BKPOW  & -- & $0.23 \pm 0.05$ & $1.70^{+0.03}_{-0.12}$ & $33.6^{+16.1}_{-28.6}$  &  4 (8)  &  8.4e-01 \\
\hline
B+X+U 4.2\,ks & POW    & $1.75\pm0.03$ & $1.04\pm0.02$ & $1.93\pm0.01$ & -  &  1372 (681)  & 6.9e-49  \\
B+X+U 4.2\,ks & BKPOW  & $1.35\pm0.03$ & $0.51 \pm 0.03$ & $1.69\pm0.01$ & $6.8\pm0.3$  &  861 (680)  &  2.7e-06 \\
\hline
B+U 43\,ks & POW    & --& $0.38\pm0.06$ & $1.89\pm0.04$ & --  &  14 (9)  & 1.2e-1  \\
B+U 43\,ks & BKPOW  & -- & $0.38\pm0.06$ & $1.39\pm0.04$ & ${7.3\times10^{-4}}^{+1.7\times 10^{-3}}_{-7.3\times 10^{-4}}$  &  14 (8)  &  8.3e-02 \\
\hline
B+X+U 43\,ks  & POW    & $1.27\pm0.03$ & $0.49\pm0.03$ & $1.83\pm0.01$ & --  &  778 (678)  &  4.2e-03 \\
B+X+U 43\,ks  & BKPOW  & $1.14\pm0.08$ & $0.28\pm0.03$ & $1.72\pm0.02$ & $4.5\pm0.2$ &  748 (677)  &  3.0e-02 \\
\enddata
\end{deluxetable*} 

\subsection{Light Curve} \label{sec:lcurve}
The combined X-ray and UV/optical light curves of \event\ are shown in Figure~\ref{fig:combined_lc} (top and bottom panels, respectively). The XRT, \maxi{}, and \nicer{} measurements are placed in a common bandpass from 0.3--10.0\,keV. The BAT survey mode data, however, are \textit{not} extrapolated to this band and instead plotted from 14-195\,keV -- due to the inference of a spectral break below this band (\S\ref{sec:SEDs}), these points are instead meant to be illustrative.

While soft X-ray observations of the afterglow did not begin until $T_{0}+2.5$\,ks, the light curve clearly lacks signatures of the canonical X-ray afterglow behavior at early times, including the prompt decay phase, the plateau phase, and any prominent flaring \citep{Nousek+2006}. This relatively simple monotonic decay has been noted previously in other GRBs detected at high ($\geq$GeV) energies \citep{Yamazaki+2020}. 

We fit the joint 0.3--10.0\,keV (XRT+\maxi{}+\nicer{}) X-ray light curve using the Markov Chain Monte Carlo software package {\sc emcee} \citep{emcee}. An additional fixed fractional error was included as a nuisance parameter to account for cross-calibration uncertainty. We find a broken power-law model is strongly preferred over a single power-law, although still formally a more complex behavior is indicated. The best-fit parameters for the broken power-law model are $\alpha_{1,x} = -1.498 \pm 0.004$, $\alpha_{2,X} = -1.672 \pm 0.008$, and $t_{\mathrm{break},X} = (7.9_{-1.0}^{+1.1}) \times 10^{4}$\,s (68\% confidence intervals). The initial \maxi{} detection at $T_{0}+2.5$\,ks clearly indicates a shallower decay at early times ($T_{0}+ \lesssim 3.3$\,ks), while a late time excess ($T_{0} + 2$\,Ms) may indicate some energy injection (e.g., \citealt{Zhang+2006}).

For the UV/optical data, we constructed a combined single-band light curve following the procedure outlined in \citet{sch10}. We then fit this light curve using the identical procedure described above for the X-ray data (including the cross-calibration nuisance parameter). We find that a single power-law and broken power-law model provide comparable quality fits to the data -- the statistically preferred model depends sensitively on how late-time ($> 10^{5}$\,ks) measurements are included in the fit. The best fit index derived for the single power-law model is $\alpha_{O} = -1.13 \pm 0.01$, while for the broken power-law model the parameters are $\alpha_{O,1} = -0.98^{+0.11}_{-0.05}$, $\alpha_{O,2} = -1.31^{+0.05}_{-0.07}$, and $t_{\mathrm{break},O} = (2.2_{-1.1}^{+1.7}) \times 10^{4}$\,s. Regardless of the model selected, the UV/optical light curve clearly declines more slowly than the X-ray emission.
 Given the large foreground extinction, these wavelengths are unaffected by contamination from an emerging supernova or an underlying host galaxy, and thus this accurately reflects the afterglow behavior.

\section{Discussion} \label{sec:disc}

\subsection{Comparison with previous GRBs} \label{sec:comp}
\begin{figure}
    \centering
    \includegraphics[width=\linewidth]{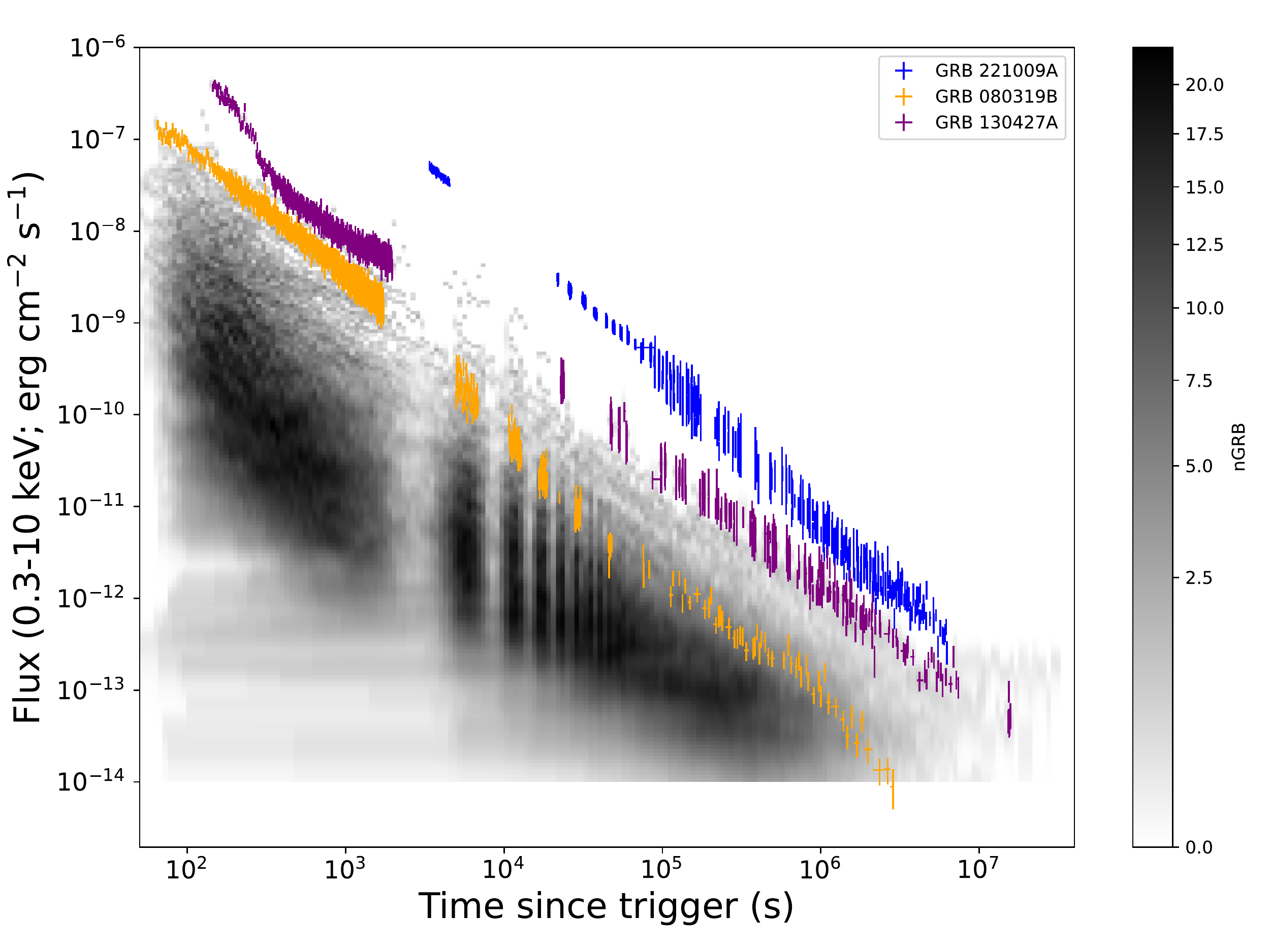}
    \includegraphics[width=\linewidth]{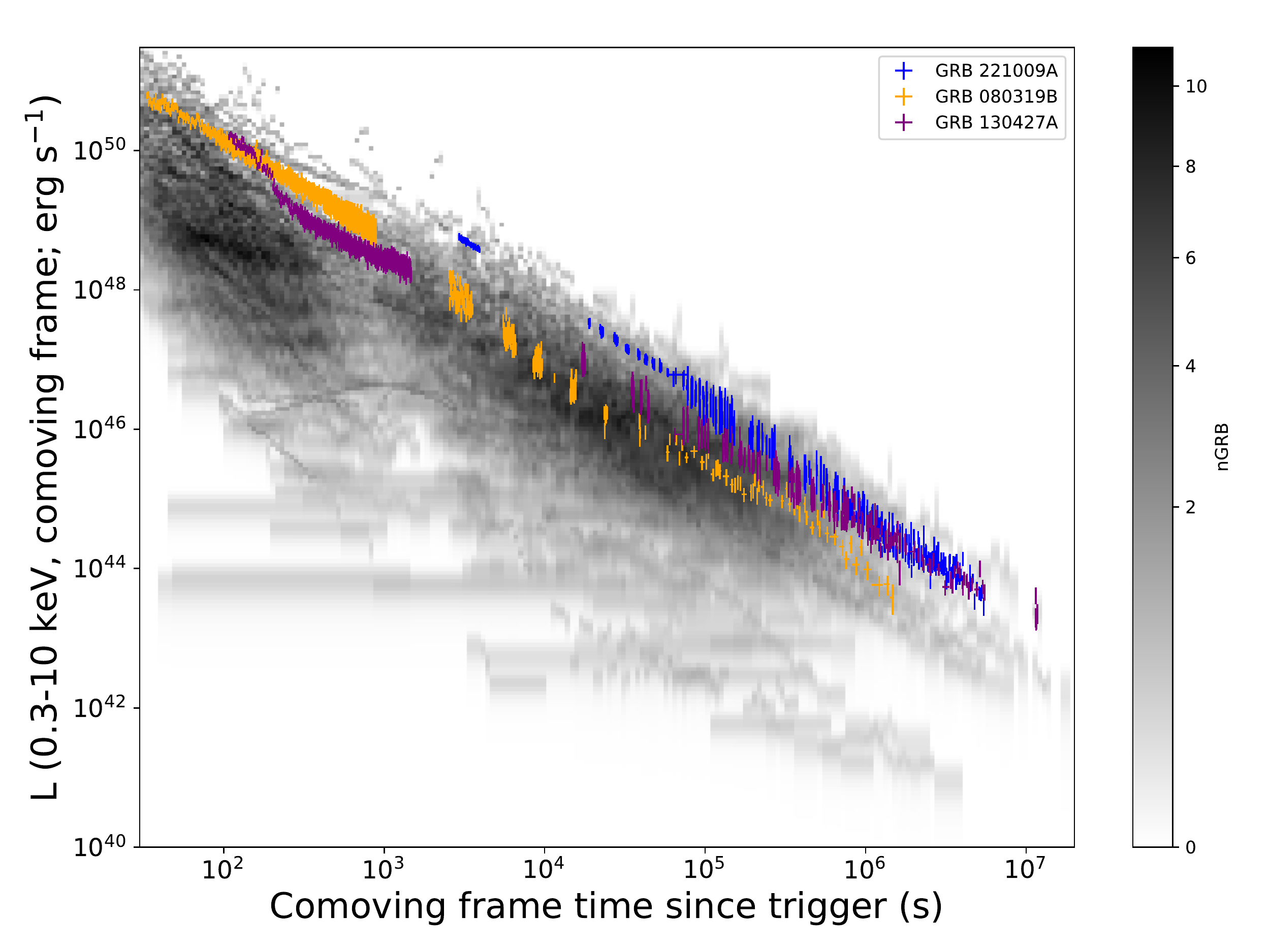}
    \caption{A comparison of \event{} with all of those observed by XRT.
    The greyscales indicate the number of GRBs in each (time, brightness) bin, the blue light curve
    is \event{}, with two notable bright GRBs, GRB\,130427A (purple) and GRB\,080319B (orange) shown for comparison. \emph{Top}: Observer frame, 0.3--10\,keV observed flux light curves. \emph{Bottom}: Comoving-frame, 0.3--10\,keV intrinsic (unabsorbed) luminosity light curves; the greyscale sample data includes only those GRBs with published redshifts.}
    \label{fig:compXRT}
\end{figure}

The top panel of Figure~\ref{fig:compXRT} plots the observed 0.3--10.0\,keV flux light curve of \event, overplotted on the entire sample of XRT light curves since \swift\ launch (all uncorrected for absorption). At the time of the first XRT observations, \event\ is approximately an order of magnitude brighter than any previous X-ray afterglow. Put differently, forming the distribution of X-ray afterglow flux at $T_{0}+4.5$\,ks (a time selected to minimize the number of events requiring interpolation due to orbital gaps in light curve coverage), \event{} falls 5.5$\sigma$ above the mean.

However, \event\ is also one of the nearest GRBs detected since the launch of \swift{} -- 12 events (out of $\approx$400) have lower redshifts reported in the online \swift{} GRB Table\footnote{See \url{https://swift.gsfc.nasa.gov/archive/grb_table}.}. To compare to the intrinsic properties of the X-ray afterglow sample, 
we converted the observed count-rate light curve into intrinsic luminosity in the 0.3--10\,keV energy band in the GRB comoving frame. The luminosity in a given bin is given by:
\begin{equation}
\label{eq:L}
L = 4\pi D_L^2 R C_u k    
\end{equation}
\noindent where $D_L$ is the luminosity distance (749.3 Mpc), $R$ is the measured 0.3--10\,keV count rate (in the observer frame), $C_u$ is the conversion from count-rate to unabsorbed 0.3--10\,keV flux (also in the observer frame) obtained from the spectral fits above (\S\ref{sec:xrt}), and $k$ is the $k$-correction of \cite{Bloom01}, which corrects from observed 0.3--10\,keV flux to the 0.3--10\,keV flux in the GRB comoving frame. To calculate $k$, we assume the spectrum is an unbroken power-law with the photon index $\Gamma=1.8$. The time axis must also be corrected to the observer frame, by dividing the time by $1+z = 1.151$. The resultant light curve is shown in the bottom panel of Figure~\ref{fig:compXRT}.

While the lack of early time observations somewhat complicates the comparison, it is clear that \event\ is at the high end of the X-ray afterglow luminosity distribution (1.8$\sigma$ above the mean). But unlike when considering this comparison in flux space, \event\ is by no means an outlier. Rather we conclude that the exceptional observed brightness of the X-ray afterglow of \event{} results from the combination of being very nearby and very intrinsically luminous: while neither the distance nor the luminosity are unprecedented, together they make \event{} unique amongst the \swift{} afterglow sample.

Comparing \event{} to the broader population of UV/optical afterglows is further complicated by the large (and uncertain) line-of-sight extinction, both foreground and in the host galaxy. With $m_{U} = 17.59_{-0.12}^{+0.13}$\,mag at $T_{0}+3574$\,s, \event{} is significantly fainter than the brightest UVOT-detected afterglows, such as GRB\,080319B ($m_{U} \approx 15.4$\,mag; \citealt{Racusin+2009,Bloom+2009}) and GRB\,130427A ($m_{U} \approx 13.9$\,mag; \citealt{Maselli+2014}) at comparable observer-frame times post-burst. But even applying only a correction for foreground (i.e., Milky Way) extinction of $A_{U} \approx 6.8$\,mag \citep{sf01}, the UV/optical emission from \event{} would likely have been the brightest observed to date had it occurred at high Galactic latitude.

If we adopt $E(B-V)_{\mathrm{host}}=0.4$\,mag and an SMC-like extinction law (\S\ref{sec:SEDs}), we infer a $U$-band host extinction of 1.8\,mag. Neglecting $k$-corrections (which are likely to be much smaller than the uncertainty in the extinction correction), we find an absolute magnitude of $M_{U} \approx -29.3$\,mag (AB). Comparing to Figure~7 in \citet{Perley+2014}, we find that the UV/optical light curve of \event{} occupies a similar location in luminosity space as the X-ray afterglow: towards the most luminous end, but entirely consistent with the existing distribution.

\subsection{Astrophysical Rate} \label{sec:rate}
Although \swift\ was behind the Earth at $T_{0}$, we use the GBM light curve and spectrum (Lesage et al. in prep.) to estimate what \event\ would have looked like to the BAT, as well as to determine the occurrence rate of such events. Adopting the time-averaged spectrum of the main pulse from $T_{0}+218.5$--277.9\,s 
(10--1000\,keV fluence of $8.293 \times 10^{-2} \ \rm erg \ cm^{-2}$), we derive a BAT ($15$--150\,keV band) fluence of $1.86 \times 10^{-2}~\ecs{}$ and an average flux of $3.13 \times 10^{-4}~\ecs{}$.
This fluence from the main pulse alone is 50$\times$ higher than the largest fluence detected to date by BAT (GRB\,130427A; \citealt{Maselli+2014}).

At $z=0.151$, the 15--150\,keV (rest-frame) isotropic prompt energy release for the main pulse is $E_{\gamma,\mathrm{iso}} = 9 \times 10^{53}$\,erg. We employ the 15--150 keV energy range to facilitate a direct comparison with the entire population of BAT GRBs with measured redshifts.

To estimate the relative occurrence rate of such an energetic event, we compare \event\ with the intrinsic long GRB luminosity function derived in \citet{Lien+2014}. We consider only the main pulse, as this dominates the burst energetics, and it is difficult to ascertain when the prompt emission ends, given the afterglow was bright enough to trigger the BAT at $T_{0}+3.4$\,ks.

We randomly generated $10^{4}$ GRBs using the intrinsic GRB rate and luminosity distribution in \citet{Lien+2014}, and each simulated burst was randomly assigned a pulse structure drawn from the real BAT GRB sample. We calculated the $E_{\gamma,\mathrm{iso}}$ of these simulated bursts, and found that only one burst had an energy release (slightly) higher than the main pulse of \event. Therefore, we conclude only $\sim$1/$10^{4}$ long GRBs are as energetic as \event. 

Using this value, we crudely estimate the rate at which such energetic GRBs are detectable by BAT:
\begin{equation}
N_{\rm BAT} \approx R_{\rm GRB} \ f_{\rm Eiso} \ f_{\rm FOV} \ f_{\rm survey} = 0.06 \ \rm yr^{-1}, 
\end{equation}
where $R_{\rm GRB}$ is the intrinsic all-sky long-GRB rate from \citet{Lien+2014}, $f_{\rm Eiso} = 1.0 \times 10^{-4}$ is the fraction of GRBs in this intrinsic sample that have $E_{\gamma,\mathrm{iso}}$ larger than \event, $f_{\rm FOV} = 1/6$ is the fraction of sky covered by the BAT field of view, and $f_{\rm survey} = 0.8$ is the fraction of time BAT is able to trigger on GRBs (neglecting spacecraft slews, SAA passage, etc.). In other words, we need to wait $\sim 1/0.06 = 17$ yr for an event like the main pulse of \event\ to occur in the BAT field of view. 

We emphasize, however, that the above estimate simply indicates that BAT should have detected $\sim$1 GRB more energetic than \event\ over its lifetime, \textit{independent of distance}. But in addition to being highly energetic, \event\ is also one of the most nearby GRBs detected by \swift. To estimate the intrinsic (volumetric) rate of comparable events, we utilize the BAT trigger simulator \citep{Lien+2014} to calculate the detectability of the prompt emission at different redshifts under different representative geometries. The full details of the simulation are provided in Appendix~\ref{sec:BATsim_appendix}.

Using this framework, we derive an upper limit on the local volumetric rate of \event-like events of  $R_{\rm GRB, comov}(z=0) \leq 6.1 \times 10^{-4} \rm \ Gpc^{-3} \ yr^{-1}$. Integrating this flat comoving rate from $z=0$ to $z=10$, we obtain an upper-limit on the all-sky rate of such events of $\leq 0.5$\,yr$^{-1}$. Comparing to the all-sky intrinsic long GRB rate of $\sim 4571 \ \rm yr^{-1}$ in \citet{Lien+2014}, the fraction of \event-like events is roughly $0.5 / 4571 \leq 1.0 \times 10^{-4}$, which is similar to the relative rate derived above. 

Even more remarkable, we can use these results to derive the rate of \event-like GRBs within the volume out to $z = 0.151$ ($1.1$\,Gpc$^{-3}$). \textbf{This implies we would need to wait over $\approx$10$^{3}$ years to detect another \event-like event within this volume.} The combination of the large energy release and the small distance make \event{} truly a once in a lifetime phenomenon.

\subsection{The Nature of Energetic GRBs} \label{sec:energetic}
Interpreting \event\ in the context of the standard afterglow synchrotron model (e.g., \citealt{sar98}) presents a number of challenges, in particular if the X-ray and optical emission is assumed to arise from a common origin (c.f., \citealt{Ghisellini+2007,MDP+2009}). To begin with, the broadband spectral fitting performed in \S\ref{sec:SEDs} strongly favors the presence of a spectral break around the soft X-ray band, with the change in spectral slope being consistent with the cooling frequency $\nu_{c}$. However, if we adopt relatively standard afterglow parameters ($\epsilon_{B} = 0.01$, $p = 2.5$, $\eta_{\gamma} = 0.15$), then we find that a cooling frequency $\approx$ 5\,keV (10$^{18}$\,Hz) at $\approx$0.1\,d post-trigger implies a jet expanding into an extremely low density circumburst environment: $n_{0} \approx 10^{-3}$\,cm$^{-3}$ for a constant-density (ISM-like) environment. Such low densities have been inferred previously for highly energetic events (e.g., \citealt{Cenko+2011}), but this remains difficult to reconcile with the massive star progenitors of long GRBs.

Furthermore, standard afterglow closure relations (e.g., \citealt{Racusin+2009}) are unable to reproduce the observed spectral and temporal power-law indices. For example, the broadband SED fits presented in \S\ref{sec:SEDs} find an optical spectral index $\beta_{O} \approx 0.7$. This is consistent with the temporal decay observed for a jet expansion into a constant-density medium ($\alpha_{0} \approx 1.1$; $p \approx 2.4$). However, this would predict a significantly shallower temporal decay in the X-rays ($\alpha_{X} \approx 1.3$) than what is observed. While expansion into a wind-like medium would predict a steeper X-ray temporal decay, the optical emission is inconsistent with this picture.

Given the exceptionally large $E_{\gamma,\mathrm{iso}}$, \event\ requires a narrow opening angle (and hence an early jet break time) in order to be compatible with the geometry-corrected energy release inferred from the pre-\swift\ sample (e.g, \citealt{Frail+2001}). Assuming an efficiency of $\approx$15\% in converting bulk kinetic energy into prompt $\gamma$-ray emission, a beaming-corrected energy release of $\leq 10^{52}$\,erg requires an opening angle of $\theta_{j} \lesssim 4^{\circ}$. For a top-hat jet expanding into a constant-density medium, this implies a jet break time of:
\begin{equation}
t_{j} \leq 3.5 \left( \frac{n_{0}}{0.1 \mathrm{cm}^{-3}} \right) ^{-1/3} \mathrm{d}
\end{equation}
where $n_{0}$ is the circumburst density in units of 0.1\,cm$^{-3}$ \citep{Sari1999}. Note that due to the -1/3 exponent, this result is relatively insensitive to the exact value of the circumburst density.

With X-ray coverage extending out to $T_{0}+70$\,d, the \swift+\maxi+\nicer\ X-ray light curve is more than adequate to search for such a geometric signature. But out to late times, the temporal decay does not steepen beyond $\alpha_{X} = 1.7$, significantly shallower than the $\alpha > 2$ behavior anticipated for a simple on-axis top-hat jet.

The X-ray afterglow steepening observed at $T_{0}+8 \times 10^{4}$\,s could conceivably be attributed to a jet break (see also \citealt{GCN:32755}). In this case, the corresponding jet would indeed be extremely narrow, with an implied opening angle of $\approx 2^{\circ}$. However, as described above, the post-break decay index is significantly shallower than expected for an on-axis top-hat jet. Interestingly, similar behavior was observed in GRB\,130427A and attributed to time-variable microphysical parameters \citep{Maselli+2014}. One alternative explanation is a complex angular structure for the jet (i.e., compared to the simple top-hat model assumed above, with a constant energy as a function of angle). The presence of energetic material outside the jet ``edges'' would naturally account for the shallower decline initially (e.g., O'Connor et al., in prep.). However, eventually when viewed sufficiently far off-axis the decay should steepen, something not observed in the X-ray light curve out to $T_{0}+70$\,d. A detailed exploration of the jet structure would require additional multi-wavelength observations, in particular radio coverage, and is thus beyond the scope of this work.

\section{Conclusions} \label{sec:conclusions}
Here we present the combined \swift{}, \maxi{}, and \nicer{} observations of \event, spanning from $T_0$ + 2.5\,ks to $T_0$ + 73 days and from UV to hard X-ray energies, providing the opportunity to study the prolonged afterglow in exquisite detail. The primary conclusions are as follows:
\begin{itemize}
\item Through dust echo tomography, we identify a series of dust clouds along the line-of-sight in our galaxy at distances ranging from 200\,pc to beyond 10\,kpc (i.e., well above the Galactic disk). As such, the dust echo from \event\ proves that echo tomography is acutely sensitive to small dust concentrations at large distances that are not easily measurable with other techniques.
\item The X-ray afterglow of \event\ is more than an order of magnitude brighter at $T_{0}+4.5$\,ks (in terms of measured flux) than any previous GRB observed by \swift. However, the intrinsic X-ray afterglow luminosity is at the high end, but consistent with, the distribution of \swift\ GRBs with redshifts. Thus, \event\ is unique \textbf{both} because it is luminous and very nearby (by GRB standards).
\item We calculate the rate of \event-like events (i.e., GRBs with prompt emission as energetic/luminous as that inferred by the GBM). We find that only $\sim$1 in $10^{4}$ GRBs has an $E_{\gamma,\mathrm{iso}}$ comparable to \event. When factoring in the distance, we find that the rate of GRBs as energetic \textbf{and} nearby as \event\ is $\lesssim$ 1 per 1000 years.
\item From the extensive multi-wavelength data sets, we find that the afterglow emission from \event\ is not well described by standard synchrotron afterglow theory. In particular, the break observed in the X-ray light curve at $T_{0}+79$\,ks is inconsistent with a jet break from an on-axis top-hat jet. Either the jet structure is more complex, with significant power outside the jet core, or \event\ is not narrowly collimated. The latter scenario would have profound implications for the energy budget of the event.
\end{itemize}

While \event\ disappeared behind the Sun for most observatories at $\approx$70\,d post-burst, the X-ray and radio afterglow are still anticipated to be easily detectable after the field becomes visible again in $\approx$February 2023. Coupled with the exquisite multi-wavelength data collected from radio to VHE gamma-rays, the broadband story of this exceptional event will continue to unfold over the coming months, years, and (perhaps) decades.

\begin{acknowledgments}
We acknowledge the use of public data from the Swift data archive. This work made use of data supplied by the UK Swift Science Data Centre at the University of Leicester. This research has made use of MAXI data provided by RIKEN, JAXA and the MAXI team.

PAE, APB, KLP, NPMK acknowledge UKSA funding.
The material is based upon work supported by NASA under award number 80GSFC21M0002. \swift{} at Penn State is supported by NASA contract NAS5-00136. 
EA, MGB, AD, PDA, AM, MP, BS, SC and GT acknowledge funding from the Italian Space Agency, contract ASI/INAF n. I/004/11/4.
PDA acknowledges support from PRIN-MIUR 2017 (grant 20179ZF5KS).
GY acknowledges the support of NASA Postdoctoral Program at the Goddard Space Flight Center, administered by Oak Ridge Associated Universities under contract with NASA.
SH acknowledges support from NSF grant AST2205917.

We acknowledge the hard work of the \swift{} and \nicer{} operations teams in performing these observations.
\end{acknowledgments}

\vspace{5mm}
\facilities{Swift (BAT, XRT, and UVOT), MAXI, NICER}

\software{                              {\sc HEASoft} \citep{heasoft};
          {\sc BatAnalysis} (Parsotan et al., in prep);
          {\sc XSPEC} \citep{XSPEC};
          {\sc astropy} \citep{astropy};
          {\sc emcee} \citep{emcee}.
          }

\appendix

\section{BAT Survey Mode and Mosaic Flux Measurements}
\label{sec:BATLC_appendix}
Details of the flux measurements and upper limits obtained from the \swift{} BAT survey mode data are shown in Table~\ref{bat_survey_table}, and the results of the daily mosaicing are presented in Table~\ref{bat_survey_mosaic_table}.

\begin{deluxetable}{ccccccc}
\tablecaption{BAT Survey Observations of \event{}. Upper limits are calculated with an assumed photon index of 1.  \newline \label{bat_survey_table}}
\tablehead{Obs ID  &	 Pointing ID	&	$T_\mathrm{start} - T_0$ &	UTC Time &	Exposure &	14--195\,keV Flux &	Photon Index \\ 
& & (ks) & & (ks) & (\ecs{}) &}
\startdata
    03111868007 &	20222810634 &	$-110.5$ & 2022 October 08 06:34:32 &	 0.661 &	$<1.11 \times 10^{-8}$ &	--- \\
                &	20222821110 &	$-7.6$ & 2022 October 09 11:10:19 &	 0.373 &	$<1.55 \times 10^{-8}$ &	--- \\
    00015314118 &	20222820117 &	$-43.2$ & 2022 October 09 01:17:08 &	 0.924 &	$<6.33 \times 10^{-9}$ &	--- \\
    00046390010 &	20222820135 &	$-42.1$ & 2022 October 09 01:34:40 &	 0.172 &	$<3.34 \times 10^{-8}$ &	--- \\
    00015314119 &	20222820626 &	$-24.6$ & 2022 October 09 06:26:19 &	   0.900 &	$<6.36 \times 10^{-9}$ &	--- \\
    00015357004 &	20222820921 &	$-14.1$ & 2022 October 09 09:21:28 &	 0.544 &	$<1.42 \times 10^{-8}$ &	--- \\
    00011105067 &	20222821046 &	$-9.1$ & 2022 October 09 10:46:08 &	 1.344 &	$<1.36 \times 10^{-8}$ &	--- \\
    00015314120 &	20222821248 &	$-1.7$ & 2022 October 09 12:48:26 &	 0.846 &	$<7.14 \times 10^{-9}$ &	--- \\
    01126854000 &	20222821422 &	3.9 & 2022 October 09 14:22:05 &	 0.627 &	$3.22^{+0.38}_{-0.37} \times 10^{-8}$ &	$2.13^{+0.19}_{-0.19}$ \\
    01126853001 &	20222821917 &	21.6 & 2022 October 09 19:17:08 &	 0.500 &	$2.61^{+0.93}_{-0.86} \times 10^{-9}$ &	$1.81^{+0.44}_{-0.44}$ \\
    01126853003 &	20222822027 &	25.8 & 2022 October 09 20:26:45 &	 0.900 &	$2.44^{+0.66}_{-0.62} \times 10^{-9}$ &	$2.31^{+0.52}_{-0.51}$ \\
    00015314121 &	20222822047 &	27.0 & 2022 October 09 20:46:57 &	 0.755 &	$3.01^{+1.11}_{-1.02} \times 10^{-9}$ &	$1.53^{+0.57}_{-0.54}$ \\
    01126853004 &	20222822156 &	31.1 & 2022 October 09 21:55:37 &	 1.744 &	$1.73^{+0.55}_{-0.5} \times 10^{-9}$ &	$1.85^{+0.49}_{-0.48}$ \\
                &	20222822332 &	36.9 & 2022 October 09 23:31:40 &	 1.500 &	$<4.09 \times 10^{-9}$ &	--- \\
                &	20222830129 &	43.9 & 2022 October 10 01:29:08 &	 1.073 &	$<4.76 \times 10^{-9}$ &	--- \\
                &	20222830243 &	48.4 & 2022 October 10 02:43:05 &	 1.742 &	$<3.82 \times 10^{-9}$ &	--- \\
                &	20222830417 &	54.0 & 2022 October 10 04:16:45 &	 1.738 &	$<3.74 \times 10^{-9}$ &	--- \\
                &	20222830754 &	67.0 & 2022 October 10 07:53:52 &	 0.914 &	$<5.05 \times 10^{-9}$ &	--- \\
                &	20222830934 &	73.0 & 2022 October 10 09:34:21 &	 0.514 &	$<6.78 \times 10^{-9}$ &	--- \\
                &	20222831234 &	83.8 & 2022 October 10 12:33:38 &	 1.314 &	$<4.61 \times 10^{-9}$ &	--- \\
                &	20222831354 &	88.6 & 2022 October 10 13:53:49 &	 1.183 &	$<4.81 \times 10^{-9}$ &	--- \\
                &	20222831531 &	94.4 & 2022 October 10 15:30:56 &	 0.900 &	$<5.34 \times 10^{-9}$ &	--- \\
                &	20222831550 &	95.6 & 2022 October 10 15:50:56 &	 0.543 &	$<6.53 \times 10^{-9}$ &	--- \\
                &	20222831704 &	100.0 & 2022 October 10 17:04:26 &	 1.679 &	$<3.76 \times 10^{-9}$ &	--- \\
    03111808004 &	20222830447 &	55.8 & 2022 October 10 04:46:38 &	 0.594 &	$<5.99 \times 10^{-9}$ &	--- \\
    00033349157 &	20222830906 &	71.3 & 2022 October 10 09:05:56 &	 1.655 &	$<3.68 \times 10^{-9}$ &	--- \\
    00015314124 &	20222831417 &	90.0 & 2022 October 10 14:17:05 &	 0.807 &	$<7.28 \times 10^{-9}$ &	--- \\
    01126853005 &	20222831837 &	105.6 & 2022 October 10 18:37:15 &	 1.500 &	$<3.80 \times 10^{-9}$ &	--- \\
    00015314125 &	20222831904 &	107.2 & 2022 October 10 19:03:48 &	 0.704 &	$<7.11 \times 10^{-9}$ &	--- \\
    01126853006 &	20222832022 &	111.9 & 2022 October 10 20:22:19 &	 1.500 &	$<4.07 \times 10^{-9}$ &	--- \\
                &	20222832147 &	117.0 & 2022 October 10 21:47:19 &	 0.900 &	$<5.17 \times 10^{-9}$ &	--- \\
                &	20222832204 &	118.2 & 2022 October 10 22:07:19 &	 0.299 &	$<8.71 \times 10^{-9}$ &	--- \\
                &	20222840108 &	129.1 & 2022 October 11 01:08:33 &	 0.839 &	$<5.31 \times 10^{-9}$ &	--- \\
                &	20222840233 &	134.2 & 2022 October 11 02:33:30 &	 0.782 &	$<5.24 \times 10^{-9}$ &	--- \\
                &	20222840444 &	142.0 & 2022 October 11 04:44:25 &	 0.362 &	$<8.01 \times 10^{-9}$ &	--- \\
                &	20222840548 &	145.8 & 2022 October 11 05:47:44 &	 1.499 &	$<3.98 \times 10^{-9}$ &	--- \\
                &	20222840735 &	152.3 & 2022 October 11 07:34:38 &	 1.500 &	$<4.04 \times 10^{-9}$ &	--- \\
                &	20222840915 &	158.3 & 2022 October 11 09:14:53 &	 1.254 &	$<4.59 \times 10^{-9}$ &	--- \\
                &	20222841047 &	163.8 & 2022 October 11 10:47:19 &	 1.453 &	$<4.26 \times 10^{-9}$ &	--- \\
    00015314126 &	20222840250 &	135.2 & 2022 October 11 02:50:06 &	 0.806 &	$<7.35 \times 10^{-9}$ &	--- \\
    00015314127 &	20222840858 &	157.3 & 2022 October 11 08:58:33 &	 0.899 &	$<6.83 \times 10^{-9}$ &	--- \\
\enddata
\end{deluxetable}

\begin{deluxetable}{ccccc}
\tablecaption{\swift{} BAT Daily and Total Mosaics of \event{}. Upper limits are calculated with an assumed photon index of 1. \label{bat_survey_mosaic_table}}
\tablehead{Start Time Bin & End Time Bin &	Exposure &	14--195\,keV Flux &	Photon Index \\ 
& & (ks) & (\ecs{}) & }
\startdata
2022 October 08 & 2022 October 09 &	0.661 &	$<2.63 \times 10^{-9}$ &	--- \\
2022 October 09 & 2022 October 10 &	6.541 &	$5.05^{+0.29}_{-0.29} \times 10^{-9}$ &	$2.30^{+0.09}_{-0.09}$ \\
2022 October 10 & 2022 October 11 &	7.710 &	$6.68^{+2.4}_{-2.1} \times 10^{-10}$ &	$2.43^{+0.61}_{-0.51}$ \\
2022 October 11 & 2022 October 12 &	4.136 &	$<6.47 \times 10^{-10}$ &	--- \\
2022 October 08 & 2022 October 12 &	19.048 &	$1.90^{+0.17}_{-0.17} \times 10^{-9}$ &	$2.28^{+0.15}_{-0.14}$ \\
\enddata
\end{deluxetable}

\section{\swift{} XRT Dust and Background Subtraction}
\label{sec:xrt_appendix}
In WT mode the \swift{} XRT CCD is read out in columns and the spatial information collapsed to one single dimension. As a result, the `source' region will contain all dust photons `above' and `below' the GRB in terms of CCD position, while the background regions will sample a different vertical slice through the dust rings, with potentially different spectral properties from those in the source region. Furthermore, the columns are summed over the full 600-pixel height of the XRT CCD, but only the central 200 columns are read out, which can make a suitable background region difficult to obtain.

\begin{figure}
\centering
\includegraphics[width=\columnwidth,trim=15 15 10 10,clip]{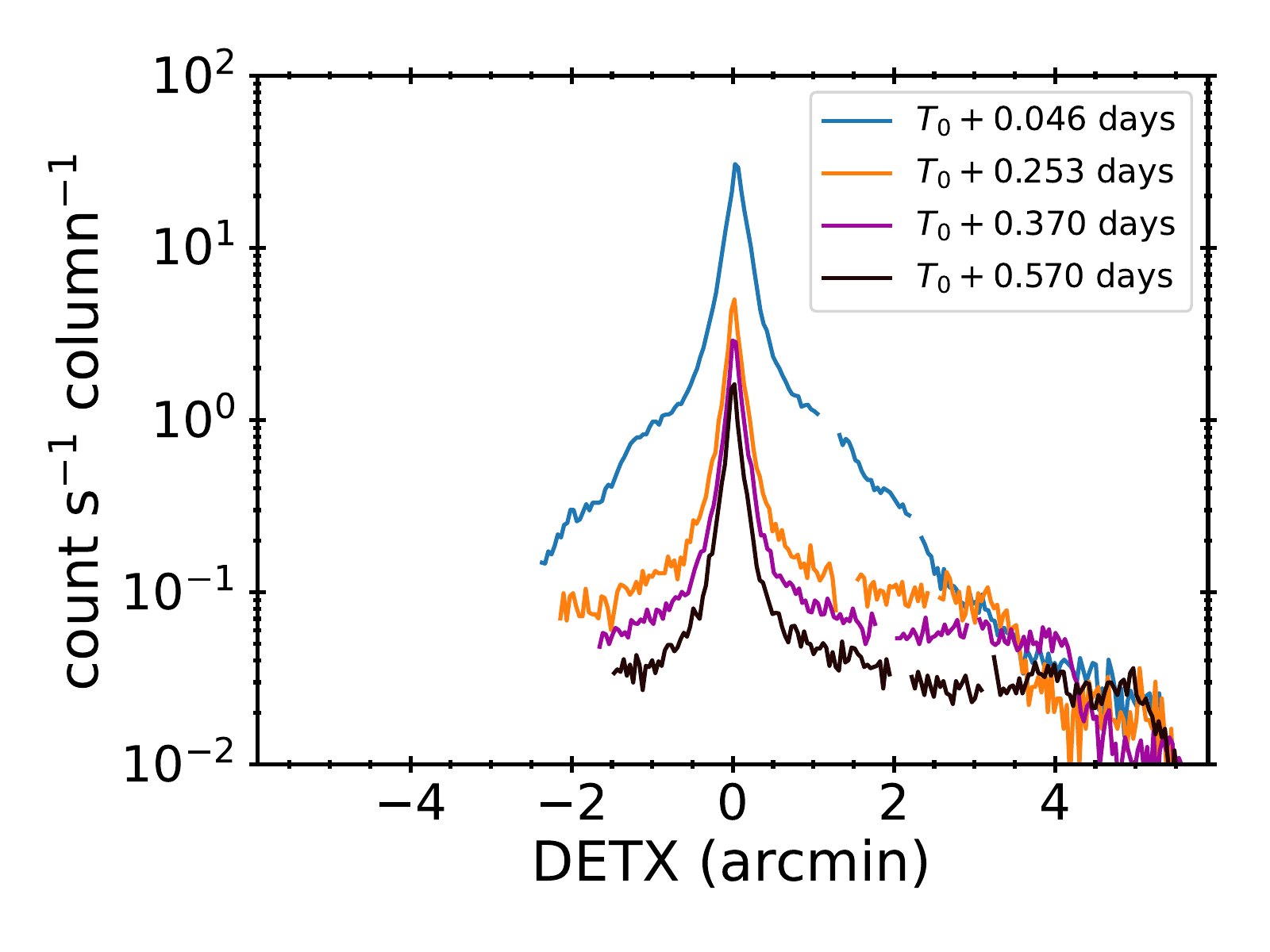}
\caption{XRT WT-mode 1-D PSF profiles as a function of time in the $0.8$--5.0\,keV band, illustrating the evolution of the echo in the early WT data.}
\label{fig:wt_profs}
\end{figure}

\begin{figure}
\centering
\includegraphics[width=\columnwidth,trim=10 10 10 0,clip]{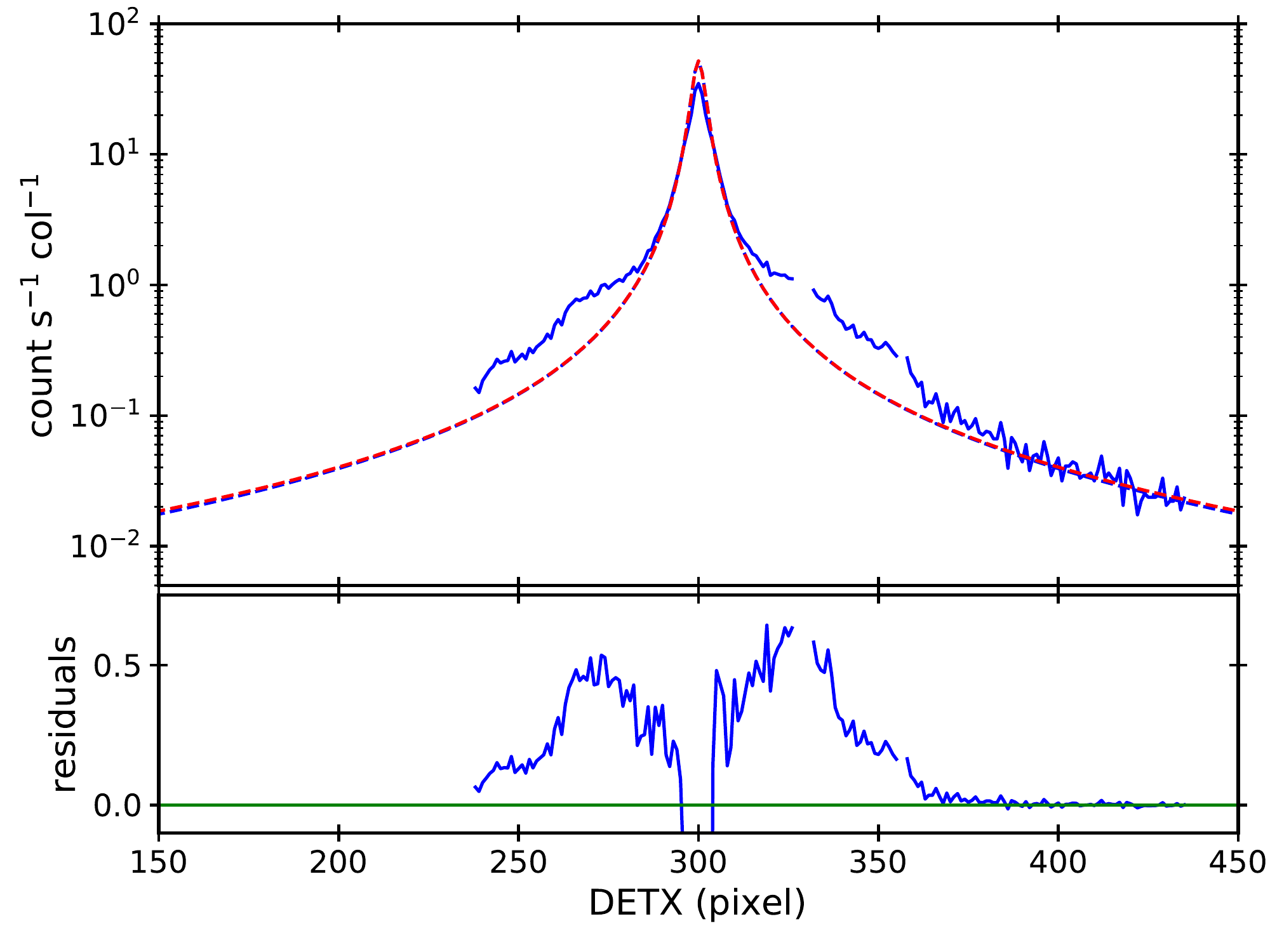}
\caption{Upper panel: XRT WT mode 1-D PSF profile from the first snapshot of data at $T_{0} + 3.4$\,ks (solid line) and the modelled point source profile (dashed line).
Lower panel: the data minus model residual. }
\label{fig:wt_snap1}
\end{figure}

Example time-sliced 1-D profiles from the WT-mode data are shown in Figure~\ref{fig:wt_profs}. These suggest that within a 20-pixel source accumulation region, the data are dominated by the source.  The profile
from the first snapshot of WT data is shown in more detail in Figure~\ref{fig:wt_snap1}, along with a fit of the profile expected from a point source; the lower panel shows the difference between the data and model. As the GRB was so bright in this snapshot, the central core profile is suppressed due to pile-up (resulting in
a negative residual, which is not shown for clarity); the central 10 columns were subsequently excluded in the spectral extraction to remove the piled-up core. The remaining residuals suggest the scattering echoes contribute $\sim 10-15$ \% of the observed count rate in a $5-20$ pixel radius extraction region centred on the source; thus the GRB spectral normalization will be overestimated by a similar amount due to the echo contamination in this snapshot.

To investigate the effect of the echo on the subsequent WT-mode flux measurements, we took the PC mode image from $T_0+1.036 - 1.111$ day and summed it in one dimension so as to mimic the WT mode profiles.  We did this twice; first using all data, then a second time after removing a model of the central source PSF, which was fit to the inner-most 20-pixel radius data. The 1-D intensity profiles thus obtained show the source-less data are approximately flat, with a variation of $<25$ \% between the source outer extraction radius and the wings out to 6 arcminutes. Within the source accumulation region, $\sim 75$ \% of the events come from the source. Thus, the source light curve is not significantly altered by the variation in the estimated background level obtained from the wings of the 1-D profile where the dust dominates.

\begin{figure}
    \centering
    \includegraphics[height=8cm]{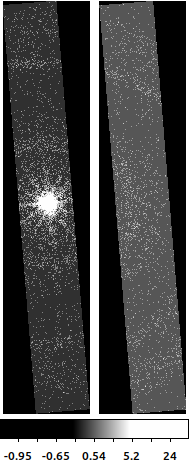}
    \caption{The extracted PC-mode data designed to mimic the WT-mode extraction to investigate 
    the impact of dust. \emph{Left}: the source region, \emph{right}: the background region. The regions are tilted at the roll angle of the spacecraft.}
    \label{fig:WTspecreg}
\end{figure}

\begin{figure}
    \centering
    \includegraphics[height=8cm]{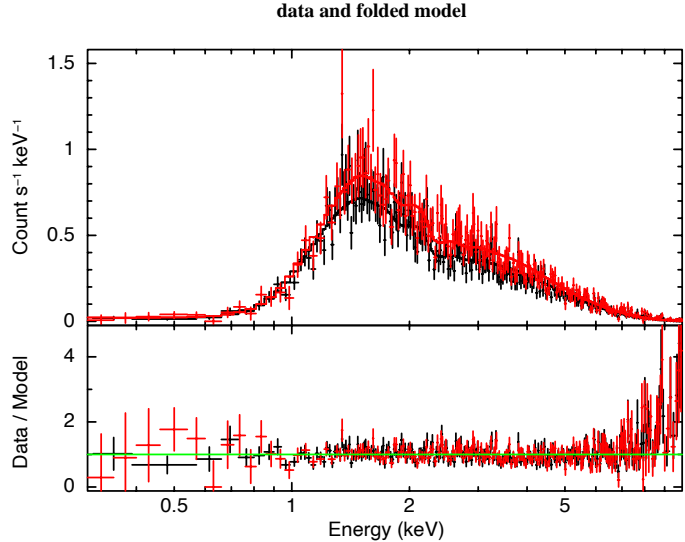}
    \caption{A comparison of a PC-mode spectrum (black), in which dust has been correctly accounted for, and a mimicked WT-mode spectrum (red) of the same data, enabling us to quantify the impact of the dust on the WT mode data.}
    \label{fig:XRTDustSpecCheck}
\end{figure}

To determine the impact of dust on the WT-mode spectroscopic data, we used the PC mode data in ObsID 01126853004, comprising 3.1\,ks of data collected between 89 and 101\,ks after the \fermi{} trigger.\footnote{Although the first snapshot from this ObsID started at 68\,ks, its ontime was only 7.5 s, meaning the exposure effectively started during the second snapshot at 89\,ks post trigger.} First we extracted data from a vertical region corresponding to a typical WT-mode background region, and also from a 20-column radius region around the source, excluding the source itself. The spectrum of the dust in the source columns is visibly slightly harder than that in the background region; fitting an absorbed power-law in {\sc XSPEC} showed the difference to be minimal, however, with the best-fitting parameters varying by less than 6\%\ between the two spectra and comfortably agreeing within their 90\%-confidence errors. To better quantify the effect of this possible change in dust spectrum between source and background regions, we extracted two spectra from the PC mode data. The first was taken from a 20-pixel radius circle centred on the source (i.e.\ with no dust present) with a background spectrum taken from the dust-free region identified above. In the second case we mimicked WT data, extracting data from the full vertical height of the CCD for regions corresponding to those used in the WT-mode analysis (Figure~\ref{fig:WTspecreg}). We fit these two spectra independently in {\sc XSPEC} using the absorbed power-law model. The results are shown in Table~\ref{tab:xrtCheckSpec} and in Figure~\ref{fig:XRTDustSpecCheck}. While all parameters agree to within their errors, the best-fit photon index is 7\%\ higher (i.e., softer) and the absorption column 27\%\ higher in the WT-style data. Since these data were gathered when the source was fainter than in the real WT data (i.e.\ the impact of dust contamination is at its worst in this experiment), we adopt these percentages as the maximum inaccuracy expected from our WT-mode spectral fits.

\begin{deluxetable}{cccc}
\tablecaption{An estimate of the impact of dust on the WT-mode spectral fit results; see text for details. \newline $^1$ (PC-WT)/PC \label{tab:xrtCheckSpec}}
\tablehead{Parameter & PC result &	Pseudo-WT result &  Percentage Difference$^1$}
\startdata
\nh\ ($10^{22}$ \cms)                             & $1.1\pm0.2$   &  $1.4\pm0.2$   &  27\% \\
Photon index                                      & $1.47\pm0.08$ &   $1.57\pm0.09$           &  7\% \\
Flux (0.3--10\,keV, $10^{-10}$ \ecs{}) & $1.64\pm0.06$ &   $1.74\pm0.06$           &  6\% \\
\enddata
\end{deluxetable}

\section{UVOT Afterglow Photometry}
\label{sec:UVOT_appendix}
The final photometry measured for \event\ is displayed in Table~\ref{uvot_photometry_table}.

\section{\maxi{} spectral analysis and dust scattering mitigation}
\label{sec:MAXIappendix}

\begin{figure}
    \centering
    \includegraphics[width=\linewidth]{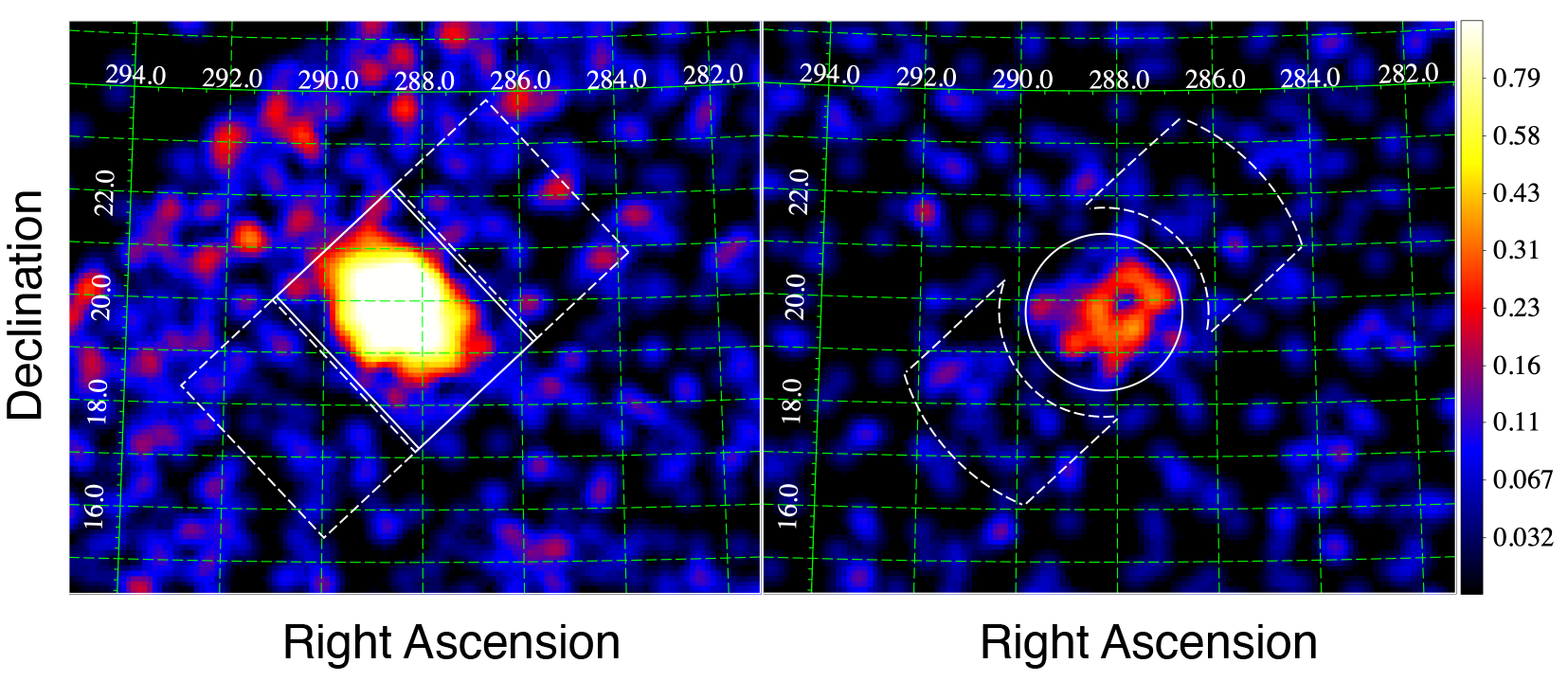}
        \caption{GSC 2--20\,keV images obtained with GSC\_5 at \tzks{+2.5} (left) and GSC\_4 at \tzks{+13.6} (right). 
    Source and background regions are shown by the solid lines and the dashed lines, respectively. }    
    \label{fig:maxi_img}
\end{figure}

Source spectra at the first (\tzks{+2.5}) and second (\tzks{+8.0}) scans were extracted from a $3.0^\circ \times 4.0^\circ$ rectangular region centered on the source (Figure~\ref{fig:maxi_img}, left),
corresponding to F{\it white}M of 1.5 deg and 1.5--2 deg of the PSF for the scan and its perpendicular (anode) 
directions, respectively \citep{sugizaki11}.
Since the source was detected near the center of the GSC\_4 detector and at the edge of the GSC\_5 detector 
($\beta$ = 2--3 deg, see Figure 2 in \citealt{mihara11}), 
we extracted the background spectra from two $2.4^\circ \times 4.0^\circ$ rectangular regions before and after scanning the source region,
avoiding a shadow region near the center frame of the GSC\_4 camera body (at $\beta \simeq 0$) and a high background region in the GSC\_5.
GSC\_4 spectra at or after the scan at 17:04 (\tzks{+13.6}) were obtained from a circular region within a radius of 1.5 arc-deg,
and background ones were from an annulus region with the inner and outer radii of 2.0 and 4.0 arc-deg
overlapped with a $8.0^\circ \times 3.4^\circ$ rectangular region (Figure~\ref{fig:maxi_img}, right).
GSC\_5 spectra at those scans were not used because of low signal-to-noise data.

We evaluated fluxes of the direct afterglow component in the following two ways:
First, we fitted the spectra in the 4--20\,keV band where the contribution from the dust scattering component is small
(middle rows in Table~\ref{tab:gscspec}).
This led to harder fitted photon indices in the first and second scan spectra, $\Gamma = 1.75\pm0.09$ and $\Gamma = 2.07^{+0.28}_{-0.26}$, than those in the simple 2-20 keV fits,
the former being closer to that of the first XRT observation, $1.61\pm0.02$.

\begin{deluxetable}{ccccccccc}
\tablecaption{\maxi{} observation logs and spectral fit results.  
\label{tab:gscspec}}
\tablehead{$T - T_0$&			&	\multicolumn{3}{c}{Flux ($10^{-8}~\ecs$)}	& &	Dust flux \\ \cline{3-5} \cline{7-7}
(ks)	&	\raisebox{1em}{Photon index}	&  0.3--10 keV		& unabs 0.3--10 keV	& 4--10 keV & & 0.3--10 keV  & \raisebox{1em}{C-stat/d.o.f}}
\startdata
		&		& \multicolumn{3}{c}{Single power-law fit (Range : 2-20 keV)}\\ \hline
2.459 	& $1.95 \pm 0.05$		& $6.98\pm0.21$		& $13.54^{+0.85}_{-0.78}$	& $3.68\pm0.11$		&& --- & 353/367\\
				& 1.85$^f$			& $6.83\pm0.19$		& $12.32\pm0.34$			& $3.77^{+0.11}_{-0.10}$	&& --- & 356/368\\
8.033	& $2.22^{+0.16}_{-0.15}$	& $1.31^{+0.10}_{-0.09}$	& $3.21^{+0.68}_{-0.51}$		& $0.60\pm0.05$		&& --- & 180/208\\
				& 1.85$^f$			& $1.22^{+0.09}_{-0.08}$	& $2.21\pm0.15$			& $0.68\pm0.05$		&& --- & 187/209\\
13.608	& 1.85$^f$			& $0.66\pm0.08$		& $1.19^{+0.15}_{-0.14}$ 		& $0.36^{+0.05}_{-0.04}$	&& --- & 93/84\\
19.181	& $1.85^f$			& $0.45\pm0.07$		& $0.80^{+0.13}_{-0.12}$		& $0.25\pm0.04$		&& --- & 59/62\\
24.757	& $1.85^f$			& $0.37^{+0.07}_{-0.06}$	& $0.66^{+0.12}_{-0.11}$		& $0.20^{+0.04}_{-0.03}$	&& --- & 40/51\\
30.330	& $1.85^f$			& $0.12\pm0.06$		& $0.22^{+0.11}_{-0.10}$		& $0.07\pm0.03$		&& --- & 54/37\\
\hline
			&	& \multicolumn{3}{c}{Single power-law fit (Range : 4-20 keV)}\\ \hline
2.459 	& $1.75\pm0.09$		& $6.01^{+0.36}_{-0.34}$	& $10.10^{+1.18}_{-1.01}$ 	& $3.46\pm0.13$		&& --- & 307/322\\
				& 1.85$^f$			& $6.34\pm0.23$	& $11.44^{+0.42}_{-0.41}$			& $3.50^{+0.13}_{-0.12}$	&& --- & 309/323\\
8.033	& $2.07^{+0.28}_{-0.26}$	& $1.17^{+0.22}_{-0.18}$  & $2.52^{+1.35}_{-0.74}$		& $0.58\pm0.06$		&& --- & 145/166\\
				& 1.85$^f$			& $1.05\pm0.10$	& $1.90^{+0.19}_{-0.18}$			& $0.58\pm0.06$		&& --- & 146/167\\
13.608	& 1.85$^f$			& $0.60\pm0.12$	& $1.08^{+0.22}_{-0.21}$			& $0.33^{+0.07}_{-0.06}$	&& --- & 42/36\\
19.181	& $1.85^f$			& $0.36^{+0.09}_{-0.08}$	& $0.65^{+0.16}_{-0.14}$		& $0.20^{+0.05}_{-0.04}$	&& --- & 43/41\\
24.757	& $1.85^f$			& $0.30\pm0.08$	& $0.55^{+0.15}_{-0.14}$			& $0.17^{+0.05}_{-0.04}$	&& --- & 29/33\\
30.330	& $1.85^f$			& $0.13^{+0.09}_{-0.08}$	& $0.24^{+0.16}_{-0.15}$		& $0.07^{+0.05}_{-0.04}$	&& --- & 52/32\\
\hline
			&	& \multicolumn{3}{c}{Two power-law fit}\\ \hline
2.459 	& $1.81 \pm 0.12$		& $6.16^{+0.63}_{-0.57}$	& $10.79^{+2.09}_{-1.66}$ & $3.47^{+0.20}_{-0.19}$	&& $1.18^{+0.81}_{-0.87}$ & 351/366\\
				&  1.85$^f$			& $6.36^{+0.28}_{-0.27}$	& $11.48^{+0.50}_{-0.49}$	 & $3.51\pm0.15$		&& $0.91^{+0.42}_{-0.41}$ & 351/367\\
8.033	& $2.14^{+0.16}_{-0.38}$	& $1.22^{+0.18}_{-0.33}$	& $2.81^{+0.92}_{-1.27}$	 & $0.59^{+0.07}_{-0.09}$	&& $0.11^{+0.48}_{-0.11}$ & 180/207\\
				&  1.85$^f$			& $0.98\pm0.13$		& $1.77\pm0.23$	 	 & $0.54\pm0.07$		&& $0.46^{+0.20}_{-0.19}$ & 181/208\\
13.608	&  1.85$^f$			& $0.53^{+0.13}_{-0.12}$	& $0.95^{+0.23}_{-0.22}$	 & $0.29\pm0.07$		&& $0.24^{+0.20}_{-0.18}$ & 92/83\\
19.181	& 1.85$^f$			& $0.27^{+0.11}_{-0.10}$	& $0.49^{+0.10}_{-0.18}$	 & $0.15^{+0.06}_{-0.05}$	&& $0.32^{+0.17}_{-0.16}$ & 55/61\\
24.757	& 1.85$^f$			& $0.25^{+0.10}_{-0.09}$	& $0.46^{+0.18}_{-0.17}$	 & $0.14\pm0.05$		&& $0.22^{+0.17}_{-0.15}$ & 38/50\\
30.330	& 1.85$^f$			& $0.12\pm0.06$		& $0.22\pm0.11$	 	 & $0.07^{+0.03}_{-0.04}$	&& $0.00\pm0.00$              & 54/36 \\
\enddata
\tablecomments{Flux columns show observed flux at 0.3--10 keV, unabsorbed flux at 0.3--10keV, and observed flux at 4--10 keV. All the exposure times are 47\,s except for 48\,s at 21:42:30. All the fluxes are in units of $10^{-8}$ \ecs. The photon index of 1.85 with no error is a fixed value. Note all errors given are $1\sigma$.}
\end{deluxetable}

\begin{figure}
    \centering
    \includegraphics[width=0.45\linewidth]{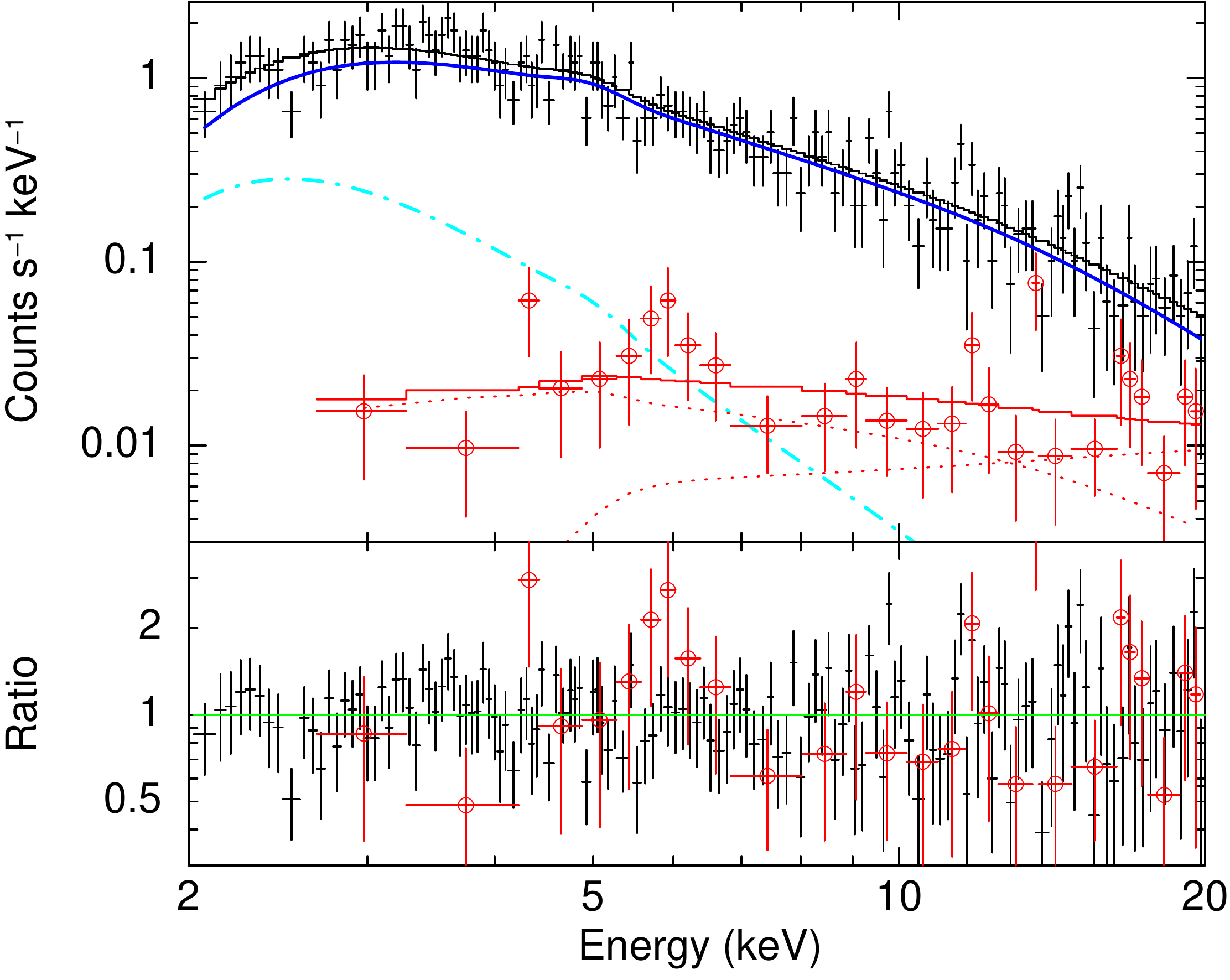}
    \includegraphics[width=0.45\linewidth]{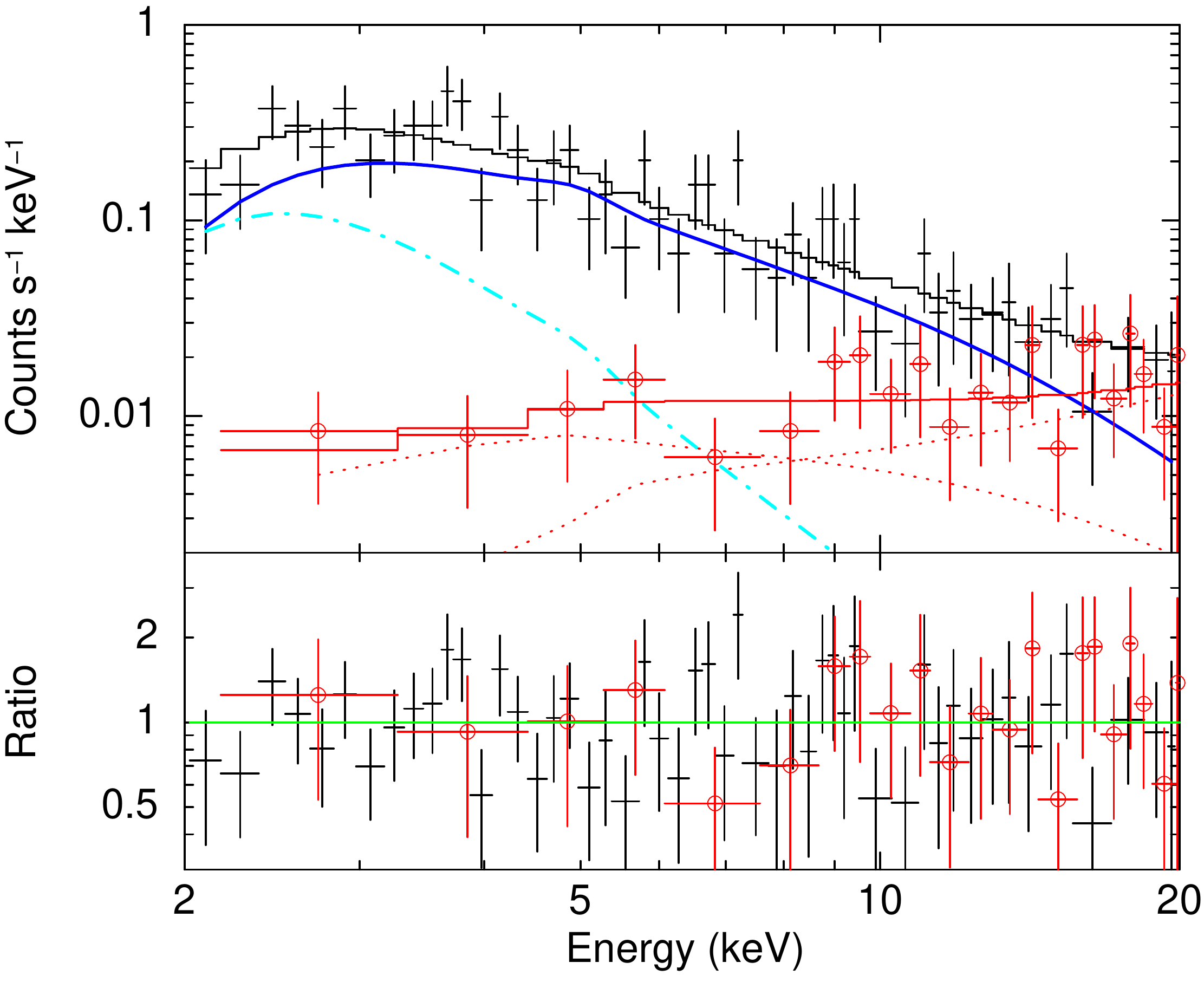}
    \caption{GSC spectra at \tzks{+2.5} (left panel) and at \tzks{+8.0} (right). 
    The blue solid lines show the direct power-law component with $\Gamma$ free for the spectrum at \tzks{+2.5}
    and fixed at 1.85 at \tzks{+8.0}. The light blue dashed lines are the power-law model with $\Gamma = 3.94$.
    The red points and lines are background data and models, respectively. The background model is the sum of two power-laws. The best-fit data to the model ratio is also shown in each panel.    }
    \label{fig:maxi_spec}
\end{figure}

Next we directly investigated energy spectra of the dust scattering component using the XRT data.
To emulate the dust scattering spectrum, we fit the XRT spectra of two bright outer dust-echo rings 
at \tzks{+95.4} and \tzks{+146.7} with the above power-law model
with the fixed column densities. 
The resultant weighted mean power-law spectral index is $\Gamma_{\mathrm{dust}} = 3.94 \pm 0.04$. 
Assuming this value does not change with time, 
we attempted to fit the GSC spectra at 2--20\,keV with a model with two power-laws, allowing the normalization of both components and the non-dust photon index for the first and second scan spectra to be free (Figure~\ref{fig:maxi_spec}).
We summarize the fitting results in the lower rows in Table~\ref{tab:gscspec}.
Interestingly, the obtained parameters for the first and second scan spectra are almost perfectly consistent with those 
in the single power-law fits at 4--20\,keV even if the photon index is fixed or not, though the uncertainty of the dust flux 
(the normalization of the power-law with $\Gamma=3.94$) is large. 
We also note that the first scan spectrum tends to be harder than the second scan one in all the cases.

The differential scattering cross-section has energy $E$ and scattering angle $\theta_\mathrm{s}$ dependence 
of approximately $\exp( - \alpha E^2 \theta_\mathrm{s}^2 )$ for $\theta_\mathrm{s} \ll 1$ where $\alpha$ is a constant of 
proportional to the square of the size of grain (e.g., \citealt{mauche86}).
Thus, an actual dust scattered component is considered to have a harder spectrum than assumed here especially in 
the first and second scan observations.
If the time and spectral evolution of the dust echo ring is understood, the fitting parameters can be more constrained. 
This is, however, beyond the scope of this paper. Finally, we note that if we assume $\Gamma = 3.5$ for the scattered power-law component
we got a steeper photon index ($\Gamma = 1.75 \pm 0.14$) and a lower 0.3--10\,keV absorbed flux of 
$5.76^{+0.82}_{-0.70}\tim{-8}$~\ecs{} for the direct power-law component in the first scan spectrum.

\section{Dust Echo Modeling}
\label{sec:dustappendix}
\begin{figure*}
    \centering
    \includegraphics[width=0.9\linewidth]{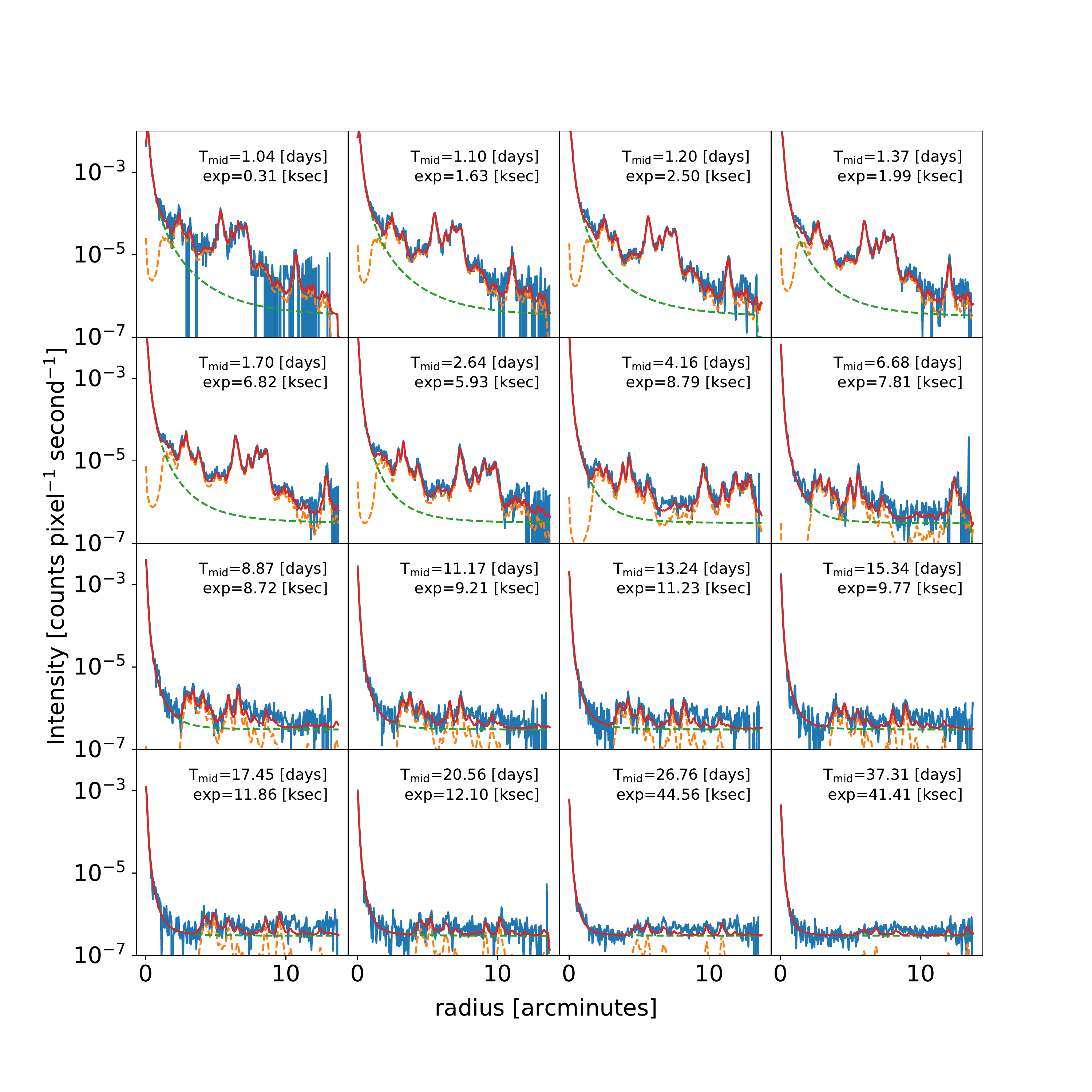}
    \caption{XRT PC mode radial intensity profiles in the 0.8--5.0\,keV band (blue solid lines); later profiles show stacked observations in indicated time windows, with radii scaled to the beginning of the observing window using eq.~(\ref{eq:dustdistance}) so that radial intensity features at the same distance appear at the same radius. Overplotted lines show the PSF and background model (green dashed) and the dust model (orange dashed) and the best fit model (solid red).}
    \label{fig:pchaloradialprofiles}
\end{figure*}

In order to examine the properties of the dust scattered echo, images were created from the XRT PC mode event data over the $0.8-5.0$\,keV band, where the dust reprocessing cross-section peaks. Given the scattered emission evolves radially with time, the images were initially extracted over per snapshot intervals (with typical exposures ranging from 0.3 to 1.6\,ks), until day 22 post trigger when per ObsID images were created (with exposures from 3.3 to 5.2\,ks). Vignetting-corrected
exposure maps were also generated, spanning the observing times for each image.

Radial profiles were then generated as follows. The GRB position was obtained on a per image basis
by fitting the expected XRT PSF model (including a piled-up profile modification when required, as described in \citealt{Evans2020}) to the imaging data out to 47.1\arcsec\ where the central source dominates. This position was then used as the location about which radial profiles were generated, initially in 2.357\arcsec\ linear bins from 0.79 to 13.75 \arcmin\ radius, following the removal of 12 faint point sources  
(with count rates less than $3.5\times 10^{-3}\thinspace{\rm count \thinspace s^{-1}}$, determined from 67\,ks of late time data ranging from $T_0 + 43$ to 64 days after the trigger). The central source profile was retained and combined with a nominal background level determined from late time data to provide the non-halo background estimate. Example radial profiles are shown in Figure~\ref{fig:pchaloradialprofiles} which shows the echo emission expanding and decreasing in intensity with time.  Later profiles were stacked to increase signal to noise in the plot, with radial bins scaled so that radial bins correspond to the first profile in the set of stacked observations according to eq.~(\ref{eq:dustdistance}) in order to align radial intensity features of different observations.

The plots show clear excess over the PSF from the afterglow emission by the point source, with each peak (i.e., each ring in the image) corresponding to a separate dust cloud along the line of sight.

Approximating the light curve of the prompt emission as a delta function, the radial intensity of the echo at time $t=t_{\rm GRB} + \Delta t$ plotted in Figure~\ref{fig:pchaloradialprofiles} can be written as
\begin{equation}
    I_{\rm echo}= \frac{2 c \cdot n(D_{\rm Dust})}{\theta^{2}} \cdot {\mathcal F} \cdot \frac{d\sigma}{d\Omega} \cdot e^{-\tau_{\rm phot}}\label{eq:dustintensity}
\end{equation}
where $n(D)$ is the dust volume density as a function of distance $D(\theta,\Delta t)$ along the line of sight for a given scattering angle $\theta$ and time since the burst $\Delta t$ and $\tau_{\rm phot}$ is the optical depth to photo-electric absorption along the line of sight. The observed radial intensity distribution is then given by the convolution of eq.~(\ref{eq:dustintensity}) with the telescope PSF.

The flux of a given cloud of column density $N$ at distance $D$ (used to perform the dist model fits) is given by \citep{heinz16}
\begin{equation}
    F=\frac{2\pi c N}{D}\cdot{\mathcal F}\cdot\frac{d\sigma}{d\Omega}
\end{equation}  

Because the time delay from the prompt emission and the radius $\theta$ of each ring are known, eq.~(\ref{eq:dustintensity}) can be used to measure the product of burst fluence, density, and scattering cross section, and, with knowledge of the fluence and a model for the scattering cross section, we can, in principle, solve for the dust density distribution $n_{\rm dust}$ as a function of line-of-sight (LOS) distance and the azimuthal angle along the ring for each observation.

In practice, estimating the fluence at soft X-ray energies (where dust scattering is most efficient and where the echo is observed) requires extrapolation from the sensitivity band of the instruments that detected the prompt emission (namely, GBM; Lesage et al. in prep.), corrected for intrinsic absorption. This fluence value, derived using GBM data (after pile-up correction procedures), has an uncertainty of $\sim$ 50\% that propagates to our results.

In addition, models for the dust scattering cross section vary from cloud to cloud and are themselves not well constrained. Thus, a conservative approach will treat both $d\sigma/d\Omega$ and $n_{\rm Dust}$ as functions to be fitted simultaneously, with the understanding that constraints derived for each will carry some degeneracy.

That said, the {\em relative} column density distribution is well constrained by this process, provided that dust chemistry and grain size distribution do not vary drastically from cloud to cloud.

At typical soft X-ray energies around 2keV, the scattering cross section is roughly constant at angles smaller than about 100\arcsec~while falling off rapidly at larger scattering angles (roughly as $d\sigma/d\Omega \propto \theta^{3.5}$). This results in a rapid decline in the intensity and flux of the echo as a function of time as the ring size expands, as can be seen in Figure~\ref{fig:pchaloradialprofiles}. It further implies that, beyond a ring radius of about 100\arcsec, smaller rings, which are farther away, will have a larger intensity per column density compared to larger rings, which are closer to the observer.

On the other hand, the Galactic latitude of \event{} of $b = 4.3^{\circ}$ suggests that the Galactic column density distribution should be dominated by nearby dust within the plane, within a few kpc. For example, dust at the intersection of the LOS with the Solar circle on the far side of the Galaxy would lie 780pc (almost 8 dust scale heights, e.g., \citealt{li18}) above the Galactic plane, where we would not expect to see substantial amounts of dust (however, any dust present at those distances will benefit from the intensity enhancement discussed in the previous paragraph).

To constrain the dust column density and cross section models, we decomposed the dust distribution along the line of sight into 100 logarithmically spaced bins and determined the least squares fit for the column density for each bin. To this end, we constructed 1-D XRT point spread functions to approximate the radial profile for each dust ring. We then defined the fitting function as given by eq.~(\ref{eq:dustintensity}) for each column density component. We added the point source PSFs for each profile to the fitting function. We then fitted all 100 components simultaneously to all the background-subtracted and radial intensity profiles between day 1 and 42 post-burst. The resulting inferred density distribution is plotted in Figure~\ref{fig:columndensityhistogram}.

\section{Broadband SED Construction}
\label{sec:SEDappendix}
The BAT survey spectra are produced using the \texttt{BatAnalysis} python package (Parsotan et al., in prep.). Using the HEASoft \texttt{batsurvey} script, the \texttt{BatAnalysis} package calculates the count rate of the GRB in each of the 8 energy bins used in the BAT survey (14--20, 20--24, 24--35, 35--50, 50--75, 75--100, 100--150, and 150--195\,keV) and the errors associated with each energy bin. The package also generates the detector response matrix using the HEASoft \texttt{batdrmgen} script for the BAT survey observation of interest. 

For the time intervals of the two SEDs, we created time-sliced X-ray spectra being careful to limit dust echo contamination (see \S\ref{sec:SEDs}).
The source spectral files were grouped to $\geq 20$ counts per energy bin. The spectral files were normalized to correspond to the 0.3--10\,keV flux of the afterglow at $T_{0}+4.2$\,ks  or $T_{0}+43$\,ks. The flux used to normalize a given spectrum is determined by fitting a power-law to the data within the SED time range and the best-fit decay index is used to compute the flux at the mid-point of the SED, in the same way as was done for the UVOT data.  

To build the optical spectral files we follow the methodology provided in \cite{sch10}. This essentially sets the count rates in the spectral files to that determined at an instantaneous epoch by extrapolating the UVOT light curves. In order to obtain the count rates at the instantaneous epoch, we first combined the data from the different filters into a single-flow light curve filter. The light curves of the different filters were normalized to that in the $v$ filter. The normalization was determined by fitting a power-law to each of the light curves in a given time range simultaneously. The power-law indices were constrained to be the same for all the filters and the normalizations were allowed to vary between the filters. The ratios of the power-law normalizations between each filter and the $v$ filter were then used to shift the individual light curves to the same normalization as the $v$ filter, thus resulting in a single filter light curve. To construct the two SEDs at $T_{0}+4.2$\,ks and $T_{0}+43$\,ks, we first determine the temporal slope from the single filter light curve within the corresponding time interval. By fixing the power-law index at this value, we then fit a power-law to the individual filter light curves. We use the derived normalizations to compute the count rate and count rate error at the required time, which was then applied to the relevant spectral file.

\section{BAT Trigger Simulation}
\label{sec:BATsim_appendix}
To estimate the intrinsic rate of such events, we utilize the BAT trigger simulator \citep{Lien+2014} to calculate the detectability of the prompt emission at different redshifts under different representative instrumental setups. Specifically, our setup uses (1) one standard average background level, (2) four different locations on the BAT image plane (i.e., Grid ID of [14, 15, 16, 17] or equivalent boresight angle of [56,45,27,0] deg), which represent different detector sensitivity and cover different locations within the BAT field of view, and (3) four different numbers of enabled detectors (29000, 22000, 18000, and 15000; these numbers represent the change of average number of enabled detectors from 2005 to 2022). We ran simulations with a combination of each of these setups with a sample of redshifts from $z = 0.1$ to $z = 12$ with increments of $\Delta z = 1.0$ (i.e., $z = 0.1, 1.0, 2.0, ...,12.0$). The simulation results are summarized in Table~\ref{tab:BAT_sim}.

\begin{table}
\begin{center}
   \caption{Detectable redshift limit for each instrumental setup. ``Grid ID'' indicates the location at the BAT image plane and corresponds to different partial coding fractions and thus different detector sensitivities. ``ndet'' refers to the number of enabled detectors.}
       \begin{tabular}{lccc}
        \hline
        \hline
       Grid ID & ndet & $z_{\rm lim}$ \\
       \hline
       17 & 29000 & 11 \\
       17 & 22000 & 11 \\
       17 & 18000 & 11 \\
       17 & 15000 & 11 \\
       16 & 29000 & 11 \\
       16 & 22000 & 11 \\
       16 & 18000 & 11 \\
       16 & 15000 & 11 \\
       15 & 29000 & 11 \\
       15 & 22000 & 10 \\
       15 & 18000 & 9 \\
       15 & 15000 & 9 \\
       14 & 29000 & 5 \\
       14 & 22000 & 4 \\
       14 & 18000 & 4 \\
       14 & 15000 & 3 \\   
       \hline
       \end{tabular}   
       \label{tab:BAT_sim}
\end{center}
\end{table}       

For each setup, we estimate the expected number of detections based on the highest detectable redshift $z_{\rm lim}$ and an assumed intrinsic rate with the following equation:
\begin{equation}
N_{\rm det, i_{setup}} =  R_{\rm GRB}(z < z_{\rm lim}) \ f_{\rm survey} \ f_{\rm fov} \ f_{\rm grid} \ f_{\rm ndet}.
\end{equation}
$R_{\rm GRB}(z < z_{\rm lim})$ is the all-sky intrinsic rate up to the redshift limit for this burst to be detected by BAT with a specific setup of Grid ID and number of enabled detectors, and is calculated by integrating the comoving rate by taking into account of the volume of the universe through the following equation (see, e.g., \citealt{Lien2011}, for details of the derivation):
\begin{equation}
R_{\rm GRB}(z < z_{\rm lim}) = 4 \pi \, \int^{z_{\rm limit}}_0 R_{\rm GRB, comov}(z=0) \, \frac{r^2_{\rm comov}}{(1+z)} \, \frac{dr_{\rm comov}}{dz} \, dz,
\end{equation}
where the comoving distance $r_{\rm comov}(z) = (\frac{C}{H_0}) \int^z_0 1/H(z) \, dz$. $f_{\rm survey}$ is the fraction of time that BAT is capable of triggering, and we adopt $f_{\rm survey}=0.8$ based on the study in \cite{Lien+2016}. $f_{\rm fov} = (2.1 \, \rm sr)/(4\pi \, \rm sr)$ is the fraction of sky covered by the entire BAT field of view down to a partial coding fraction of $\sim 0.1$,  and $f_{\rm grid} = [0.460, 0.346, 0.109, 0.085]$ is the fraction of BAT field of view for Grid ID 14, 15, 16, and 17 \footnote{\url{https://asd.gsfc.nasa.gov/Craig.Markwardt/bat-cal/solid-angle/}}. In other words, $f_{\rm fov} \times f_{\rm grid}$ is the fraction of bursts in the entire sky that would have the partial coding fraction of the specific Grid ID. $f_{\rm ndet}$ is the fraction of GRBs in the BAT field of view that would occur with this number of enabled detectors. For simplicity, we assume an equal number of GRBs occur with these four numbers of enabled detectors. That is, $f_{\rm ndet} = 0.25$.

The total number of detections can then be calculated by adding up $N_{\rm det, i_{setup}}$ for all 16 combinations of instrumental setups listed in Table~\ref{tab:BAT_sim}. That is,
\begin{equation}
N_{\rm det, tot} = \sum^{16}_{\rm i_{setup} = 1} N_{\rm det, i_{setup}}
\end{equation}

We assume a flat intrinsic comoving rate of $R_{\rm GRB, comov}(z=0)$ and adjust the value until the detection rate matches with that of a \event{}-like event. We set the upper limit of the BAT detection rate of such events to be 1 per 18 years of the \swift{} mission lifetime, because the prompt emission of the burst actually occurred outside of the BAT field of view. This gives us a corresponding upper limit of $R_{\rm GRB, comov}(z=0) \leq 6.1 \times 10^{-4} \rm \ Gpc^{-3} \ yr^{-1}$. 
Integrating this flat comoving rate from $z=0$ to $z=12$, we obtain an upper-limit on the all-sky rate of \event{}-like events to be $0.5 \ \rm yr^{-1}$. Comparing to the all-sky intrinsic long-GRB rate of $\sim 4571 \ \rm yr^{-1}$ in \cite{Lien+2014}, the fraction of \event{}-like events is roughly $0.5 / 4571 \leq 1.0 \times 10^{-4}$. 

\startlongtable
\begin{deluxetable*}{ccccccccc}
\tablecaption{\swift{} UVOT observations. 
\label{uvot_photometry_table}}
\tablehead{
\colhead{$T_\mathrm{mid} - T_0$} & \colhead{Half Exposure} & \colhead{Magnitude} &  \colhead{Flux}  & \colhead{Filter} & \colhead{S/N} \\
(ks) & (s) &  &   (\ecs{}~{\AA}$^{-1}$) &  &   
}
\startdata 
\hline
   3.4 & 25 & $16.743^{+0.051}_{-0.048}$ & $9.64\pm0.44 \times 10^{-16}$ & {\it white} & 21.956 \\
   3.5 & 30 & $16.762^{+0.046}_{-0.045}$ & $9.48\pm0.40 \times 10^{-16}$ & {\it white} & 23.877 \\
   3.5 & 20 & $16.726^{+0.057}_{-0.054}$ & $9.80\pm0.50 \times 10^{-16}$ & {\it white} & 19.698 \\
   3.8 & 10 & $16.71^{+0.097}_{-0.089}$ & $9.94\pm0.85 \times 10^{-16}$ & {\it white} & 11.664 \\
   4.0 & 10 & $16.958^{+0.103}_{-0.094}$ & $7.91\pm0.72 \times 10^{-16}$ & {\it white} & 11.024 \\
   4.1 & 75 & $16.972^{+0.075}_{-0.07}$ & $7.81\pm0.52 \times 10^{-16}$ & {\it white} & 15.078 \\
   4.4 & 10 & $16.976^{+0.104}_{-0.095}$ & $7.78\pm0.71 \times 10^{-16}$ & {\it white} & 10.976 \\
  21.9 & 246 & $18.803^{+0.063}_{-0.06}$ & $1.45\pm0.08 \times 10^{-16}$ & {\it white} & 17.686 \\
  44.9 & 76 & $19.862^{+0.166}_{-0.144}$ & $5.45\pm0.77 \times 10^{-17}$ & {\it white} & 7.065 \\
  61.2 & 408 & $20.1^{+0.085}_{-0.079}$ & $4.38\pm0.33 \times 10^{-17}$ & {\it white} & 13.261 \\
 120.0 & 14394 & $21.152^{+0.079}_{-0.074}$ & $1.66\pm0.12 \times 10^{-17}$ & {\it white} & 14.176 \\
 152.3 & 17859 & $21.655^{+0.099}_{-0.09}$ & $1.05\pm0.09 \times 10^{-17}$ & {\it white} & 11.522 \\
 198.5 & 23982 & $21.599^{+0.21}_{-0.176}$ & $1.10\pm0.19 \times 10^{-17}$ & {\it white} & 5.685 \\
 258.6 & 31606 & $21.987^{+0.19}_{-0.162}$ & $7.71\pm1.24 \times 10^{-18}$ & {\it white} & 6.218 \\
 306.9 & 6602 & $22.574^{+0.324}_{-0.249}$ & $4.49\pm1.16 \times 10^{-18}$ & {\it white} & 3.872 \\
 436.1 & 49190 & $22.703^{+0.486}_{-0.334}$ & $3.99\pm1.44 \times 10^{-18}$ & {\it white} & 2.773 \\
 585.3 & 71725 & $22.688^{+0.543}_{-0.36}$ & $4.04\pm1.59 \times 10^{-18}$ & {\it white} & 2.541 \\
 754.3 & 91946 & $22.594^{+0.573}_{-0.373}$ & $4.41\pm1.81 \times 10^{-18}$ & {\it white} & 2.440 \\
971.9 & 120245 & $>22.625$ & $<$$4.28 \times 10^{-18}$ & {\it white} & --- \\
1251.6 & 154713 & $>22.882$ & $<$$3.38 \times 10^{-18}$ & {\it white} & --- \\
1701.2 & 192141 & $>22.836$ & $<$$3.52 \times 10^{-18}$ & {\it white} & --- \\
2087.8 & 137716 & $>23.166$ & $<$$2.60 \times 10^{-18}$ & {\it white} & --- \\
3940.8 & 432429 & $>23.602$ & $<$$1.74 \times 10^{-18}$ & {\it white} & --- \\
\hline
   3.4 & 5 & $15.463^{+0.162}_{-0.141}$ & $2.44\pm0.34 \times 10^{-15}$ & {\it v} & 7.210 \\
   3.9 & 10 & $15.538^{+0.125}_{-0.112}$ & $2.28\pm0.25 \times 10^{-15}$ & {\it v} & 9.220 \\
   4.2 & 114 & $15.729^{+0.094}_{-0.086}$ & $1.91\pm0.16 \times 10^{-15}$ & {\it v} & 12.072 \\
   4.4 & 10 & $15.893^{+0.145}_{-0.128}$ & $1.64\pm0.21 \times 10^{-15}$ & {\it v} & 7.995 \\
  38.2 & 412 & $18.152^{+0.077}_{-0.072}$ & $2.05\pm0.14 \times 10^{-16}$ & {\it v} & 14.642 \\
  55.3 & 412 & $18.755^{+0.11}_{-0.1}$ & $1.18\pm0.11 \times 10^{-16}$ & {\it v} & 10.375 \\
  92.8 & 8959 & $19.427^{+0.123}_{-0.11}$ & $6.34\pm0.68 \times 10^{-17}$ & {\it v} & 9.342 \\
436.6 & 49188 & $>20.443$ & $<$$2.49 \times 10^{-17}$ & {\it v} & --- \\
 573.7 & 59918 & $21.082^{+0.698}_{-0.421}$ & $1.38\pm0.66 \times 10^{-17}$ & {\it v} & 2.109 \\
822.1 & 97374 & $>20.64$ & $<$$2.08 \times 10^{-17}$ & {\it v} & --- \\
1065.7 & 122800 & $>21.411$ & $<$$1.02 \times 10^{-17}$ & {\it v} & --- \\
1372.4 & 160437 & $>21.617$ & $<$$8.44 \times 10^{-18}$ & {\it v} & --- \\
1833.6 & 220177 & $>22.032$ & $<$$5.76 \times 10^{-18}$ & {\it v} & --- \\
2151.2 & 74786 & $>21.125$ & $<$$1.33 \times 10^{-17}$ & {\it v} & --- \\
3940.0 & 432472 & $>21.862$ & $<$$6.74 \times 10^{-18}$ & {\it v} & --- \\
\hline
   3.8 & 10 & $17.062^{+0.144}_{-0.127}$ & $9.70\pm1.20 \times 10^{-16}$ & {\it b} & 8.058 \\
   4.0 & 10 & $17.227^{+0.155}_{-0.136}$ & $8.34\pm1.11 \times 10^{-16}$ & {\it b} & 7.495 \\
   4.4 & 92 & $17.551^{+0.144}_{-0.127}$ & $6.18\pm0.77 \times 10^{-16}$ & {\it b} & 8.073 \\
  44.4 & 454 & $20.268^{+0.145}_{-0.128}$ & $5.06\pm0.63 \times 10^{-17}$ & {\it b} & 7.999 \\
  60.3 & 453 & $20.666^{+0.237}_{-0.194}$ & $3.51\pm0.69 \times 10^{-17}$ & {\it b} & 5.100 \\
438.9 & 52551 & $>21.71$ & $<$$1.34 \times 10^{-17}$ & {\it b} & --- \\
584.9 & 71610 & $>21.597$ & $<$$1.49 \times 10^{-17}$ & {\it b} & --- \\
 753.8 & 92017 & $22.444^{+0.512}_{-0.346}$ & $6.82\pm2.56 \times 10^{-18}$ & {\it b} & 2.662 \\
971.7 & 120321 & $>22.02$ & $<$$1.01 \times 10^{-17}$ & {\it b} & --- \\
1251.2 & 154869 & $>22.124$ & $<$$9.16 \times 10^{-18}$ & {\it b} & --- \\
1700.9 & 192324 & $>22.229$ & $<$$8.32 \times 10^{-18}$ & {\it b} & --- \\
2087.2 & 137713 & $>22.48$ & $<$$6.60 \times 10^{-18}$ & {\it b} & --- \\
3849.4 & 438307 & $>22.969$ & $<$$4.21 \times 10^{-18}$ & {\it b} & --- \\
\hline
   3.6 & 38 & $17.592^{+0.129}_{-0.115}$ & $3.24\pm0.36 \times 10^{-16}$ & {\it u} & 8.927 \\
   3.7 & 50 & $17.673^{+0.115}_{-0.104}$ & $3.01\pm0.30 \times 10^{-16}$ & {\it u} & 9.933 \\
   3.7 & 37 & $17.629^{+0.131}_{-0.117}$ & $3.13\pm0.36 \times 10^{-16}$ & {\it u} & 8.773 \\
   3.9 & 10 & $17.882^{+0.309}_{-0.24}$ & $2.48\pm0.61 \times 10^{-16}$ & {\it u} & 4.041 \\
   4.4 & 97 & $17.765^{+0.196}_{-0.166}$ & $2.77\pm0.46 \times 10^{-16}$ & {\it u} & 6.043 \\
  32.4 & 415 & $20.258^{+0.207}_{-0.174}$ & $2.78\pm0.48 \times 10^{-17}$ & {\it u} & 5.769 \\
  49.7 & 414 & $20.729^{+0.313}_{-0.243}$ & $1.80\pm0.45 \times 10^{-17}$ & {\it u} & 3.992 \\
  95.8 & 414 & $21.106^{+0.438}_{-0.311}$ & $1.27\pm0.42 \times 10^{-17}$ & {\it u} & 3.010 \\
438.8 & 52496 & $>20.903$ & $<$$1.54 \times 10^{-17}$ & {\it u} & --- \\
584.7 & 71385 & $>20.647$ & $<$$1.94 \times 10^{-17}$ & {\it u} & --- \\
753.5 & 91837 & $>21.678$ & $<$$7.52 \times 10^{-18}$ & {\it u} & --- \\
971.4 & 120396 & $>21.285$ & $<$$1.08 \times 10^{-17}$ & {\it u} & --- \\
1251.1 & 154783 & $>21.32$ & $<$$1.05 \times 10^{-17}$ & {\it u} & --- \\
1700.8 & 192285 & $>21.335$ & $<$$1.03 \times 10^{-17}$ & {\it u} & --- \\
2086.9 & 137473 & $>20.098$ & $<$$3.23 \times 10^{-17}$ & {\it u} & --- \\
3848.7 & 438428 & $>22.798$ & $<$$2.68 \times 10^{-18}$ & {\it u} & --- \\
\hline
   3.9 & 10 & $18.238^{+0.671}_{-0.412}$ & $2.02\pm0.93 \times 10^{-16}$ & {\it uvw1} & 2.168 \\
4.4 & 97 & $>19.108$ & $<$$9.06 \times 10^{-17}$ & {\it uvw1} & --- \\
31.6 & 450 & $>20.175$ & $<$$3.39 \times 10^{-17}$ & {\it uvw1} & --- \\
48.8 & 450 & $>20.003$ & $<$$3.97 \times 10^{-17}$ & {\it uvw1} & --- \\
67.5 & 450 & $>20.5$ & $<$$2.51 \times 10^{-17}$ & {\it uvw1} & --- \\
92.4 & 2910 & $>21.471$ & $<$$1.03 \times 10^{-17}$ & {\it uvw1} & --- \\
685.4 & 269 & $>20.833$ & $<$$1.85 \times 10^{-17}$ & {\it uvw1} & --- \\
1458.7 & 22863 & $>22.436$ & $<$$4.23 \times 10^{-18}$ & {\it uvw1} & --- \\
\hline
3.9 & 10 & $>18.228$ & $<$$2.37 \times 10^{-16}$ & {\it uvm2} & --- \\
4.4 & 97 & $>18.771$ & $<$$1.44 \times 10^{-16}$ & {\it uvm2} & --- \\
87.1 & 2398 & $>21.314$ & $<$$1.38 \times 10^{-17}$ & {\it uvm2} & --- \\
\hline
3.9 & 10 & $>18.852$ & $<$$1.54 \times 10^{-16}$ & {\it uvw2} & --- \\
4.1 & 114 & $>19.356$ & $<$$9.68 \times 10^{-17}$ & {\it uvw2} & --- \\
4.4 & 10 & $>18.966$ & $<$$1.39 \times 10^{-16}$ & {\it uvw2} & --- \\
25.6 & 83 & $>19.614$ & $<$$7.64 \times 10^{-17}$ & {\it uvw2} & --- \\
37.3 & 450 & $>20.917$ & $<$$2.30 \times 10^{-17}$ & {\it uvw2} & --- \\
54.4 & 450 & $>20.68$ & $<$$2.86 \times 10^{-17}$ & {\it uvw2} & --- \\
73.3 & 253 & $>21.153$ & $<$$1.85 \times 10^{-17}$ & {\it uvw2} & --- \\
100.5 & 450 & $>20.525$ & $<$$3.30 \times 10^{-17}$ & {\it uvw2} & --- \\
\hline
\enddata
\tablecomments{Columns 1 and 2 give the mid-time of the exposure in kiloseconds since the GBM trigger and the length of the exposure divided by 2. Magnitudes are given in the Vega system. Uncertainties are given at the  $1\sigma$ level. Observations with S/N greater than or equal to 2 were considered detections. For non-detections with S/N less than 2, $3\sigma$ upper limits are given. The values in this table have not been corrected for Galactic extinction.}
\end{deluxetable*}

\bibliography{main}{}
\bibliographystyle{aasjournal}

\end{document}